\documentclass[]{aa}
\usepackage[varg]{txfonts}
\usepackage{graphicx}
\usepackage{booktabs}
\usepackage{amsmath}
\usepackage{colortbl}
\usepackage{subfig}
\usepackage{rotating}
\usepackage{amssymb}
\usepackage{latexsym}
\usepackage{ifthen}
\usepackage{enumitem}
\def\lsim{\lower.5ex\hbox{$\; \buildrel < \over \sim \;$}}
\def\gsim{\lower.5ex\hbox{$\; \buildrel > \over \sim \;$}}
\def\be{\begin{equation}}
\def\bea{\begin{eqnarray}}
\def\eea{\end{eqnarray}}
\def\ee{\end{equation}}

\def\mdtj{{\dot {\cal M}}}
\def\mdtjc{{\dot {\cal M}}_c}

\def\bc{\begin{center}}
\def\ec{\end{center}}

\def\etal{{\em et al.}}
\def\ie{{\em i.e.,}}
\def\ep{{e^--p^+}}
\def\el{{e^--e^+}}

\def\mbh{M_{\rm B}}
\def\rg{r_{\rm g}}
\def\rs{r_{\rm s}}

\def\tg{t_{\rm g}}
\def\xsh{x_{\rm sh}}

\def\ep{{{\rm e}^--{\rm p}^+}}

\def\gamt{\gamma_{\rm \small T}}
\def\vt{v_{\rm \small T}}
\def\veq{v_{\rm eq}}

\def\od{\rm \small D}
\def\msol{M_\odot}
\def\frad{{\cal F}_{\rm rd}}

\begin{document}

\title{Radiatively driven relativistic jets in Schwarzschild space-time} 

\author{Mukesh K. Vyas
		\inst{\ref{i1},2}\thanks{E-mail: mukesh.vyas@aries.res.in}
		\and 
	Indranil Chattopadhyay
		\inst{\ref{i1}}\thanks{E-mail: indra@aries.res.in}	
		}

\institute{$^1$Aryabhatta Research Institute of Observational Sciences 
(ARIES), Manora Peak, Nainital-263002, India\\\label{i1}
$^2$Dept of Physics and Astrophysics, Delhi University, India.
}

\date{Received -- / Accepted }

\abstract {} 
{
-- 
We carry out a general relativistic study of radiatively driven, conical fluid jets around non-rotating black holes and investigate the effects and significance of radiative acceleration, as well as radiation drag.}
{
--
We apply relativistic equations of motion in curved space-time around a Schwarzschild black hole for axis-symmetric 1-D jet in steady state, plying through the radiation field of the accretion disc. Radiative moments are computed using information of
curved space-time. Slopes of physical variables at the sonic points are found using L$^{\prime}$H\^opital's rule and employed Runge-Kutta's $4^{th}$ order method to solve equations of motion. The analysis is carried out, using the relativistic equation of state of the jet fluid.}
{
--
 The terminal speed of the jet depends on how much thermal energy is converted into jet momentum
 and how much radiation momentum is deposited on to the jet. 
 Many classes of jet solutions with single sonic points, multiple sonic points as well as,
those having radiation driven internal shocks are obtained. Variation of all flow variables along the jet-axis has been studied.
Highly energetic electron-proton jets can be accelerated by intense radiation to terminal Lorentz factors $\gamt\sim 3$. Moderate terminal speed $v_{\rm \small T} \sim 0.5$ is obtained for moderately luminous discs. Lepton dominated jets may achieve $\gamt \sim 10$.
} 
{
-- Thermal driving of the jet itself and radiation driving by accretion disc photons produce a wide-ranging jet solutions staring from moderately strong jets to the relativistic ones. Interplay of intensity
and nature of radiation field and the energetics of the jet result in such a variety in jet solutions.
We show that radiation field is able to induce steady shocks in jets, one of the criteria to explain high energy power law emission observed in spectra of some of the astrophysical objects.}

\maketitle
\titlerunning{Radiatively and thermally driven jets in curved spacetime} 

\section[Introduction]{Introduction}
\label{sec1}
While analysing an optical image of M87, \cite{c18} made a note : ``curious straight ray...connected with the nucleus", which was later identified and termed as `relativistic jet' \citep{bm54}.
Since then, the observational study of jets has been revolutionized and
established as ubiquitous astrophysical phenomena associated with various classes of objects like active galactic
nuclei (AGN e.g., M87), young stellar 
objects (YSO e.g., HH 30, HH 34), X-ray binaries (e. g., SS433, Cyg
X-3, GRS 1915+105, GRO 1655-40), Gamma ray bursts (e. g., GRB 980519), Pulsar
Wind Nebulae \citep{pbok17} etc. 

This paper  investigates the properties of relativistic jets around black hole (hereafter BH) candidates like
X-ray binaries and AGNs. In such systems, jets can only emerge from accreting matter, as BHs neither have hard surface nor they are capable of emission. This fact is supported by strong correlation observed between spectral state of the accretion disc and jet. \citep{gfp03,fgr10,rsfp10}. Observations also limit the jet generation region to a distance less than $100$ Schwarzschild radii ($r_{\rm s}$) around the central object \citep{jbl99,detal12}. This implies that the entire accretion disc does not take part in jet generation.

Ever since the emergence of the first theoretical model of accretion discs i. e., the Keplerian disc or KD \citep{ss73}, or later disc models like the thick disc or TD \citep{pw80}, the advection dominated accretion flow or ADAF \citep{nkh97} and advective discs
\citep{f87,c89}, there have been many attempts to understand how photons radiated
from these discs interact with jets emerging from them. The equations of motion (EoM) of radiation hydrodynamics (RHD)
were developed 
by many authors \citep{hs76,mm84,kfm98} in special relativity (SR). Later the general relativistic (GR) version of those equations
was also obtained \citep{p06, t07}. Many authors used these EoMs under a variety of approximations
to study radiatively driven jets. \cite{i80} studied the matter flow in the radiation field above a Keplerian disc. \citet{sw81} studied particle jets in SR regime, driven by the radiation field in the funnel of a
thick accretion disc and obtained terminal speed
$\vt \sim 0.4c$ ($c{\equiv}$speed of light in vacuum) for electron-proton or $\ep$ jets, although the
terminal Lorentz factor obtained was $\gamt \sim 3$ for electron-positron or $\el$ jets. 
\citet{i89} obtained a theoretical upper limit or `magic speed' $v_{\rm m}=0.45c$ above a KD using the near disc approximation for radiation field. Any speed above $v_{\rm m}$ would
invoke radiative deceleration induced by radiation drag. Around the same time, Ferrari et. al. (1985, hereafter
FTRT85) studied
radiation interaction with a fluid jet in SR regime. They mostly assumed isothermal jets
with non-radial cross-section. A Newtonian gravitational field was added ad hoc to the EoM.
The radiation field was computed from
disc models for a variety of disc thickness. They obtained mildly relativistic jets and shocks induced by the non radial nature of the jet cross-section, as well as the radiation field.
\cite{f96} studied radiatively driven off-axis particle jets, using the radiation field similar to Icke. The detailed radiation field around BH was calculated by \cite{hf01} above a KD governed by a point mass gravity
using Newtonian and pseudo-Newtonian potentials (pNp) to mimic non-rotating and rotating BH exterior. The strength of the radiation field using Schwarzschild pNp was found to be half of Newtonian potential, but it was about one order greater for Kerr pNp. In another attempt, \citet{fth01} considered a hybrid disc
consisting of outer KD and inner ADAF. Such a scenario did produce jets with $\gamt \sim 2$, and also induced collimation.

It may be noted that a large number of jet studies in recent years have been in the realm of numerical simulations. Most of these works investigate how special relativistic jets interact with
the ambient medium, or how magnetic field affects them \citep{dh94,mm97,agmimaah01,kbvk07,mrb10}.
\cite{tnm11}, on the other hand, simulated magnetically arrested disc and jet launching from such a disc.  Although not a simulation, but \cite{msvtt06} studied steady jets in the meridional plane in general relativistic magneto hydrodynamics (GRMHD). These kind of studies are important because they enhance
the understanding of the system as well as, acts as test cases for numerical simulations.

Most of the jet simulations did not include radiatively driven jets.
Simulations which did include interaction of radiation with outflows were mainly in the non-relativistic limit
\citep{cc02b,csnr12}. There is a general consensus that radiation cannot accelerate fluid jets to relativistic speeds \citep{ggmm02} and probably that is the reason, why simulations on radiation driving of jets
are few in number. General relativistic simulations which includes the interaction of radiation with matter
are there, but they studied
either the stellar collapse scenario \citep{flls08}, or Bondi-Hoyle accretion \citep{zrrd11} and that too
in optically thick medium. Jets are divergent flow and are optically thin.

%So the present effort is not
%only important to assess the applicability of radiation as accelerating agent and how the anisotropic
%radiation field affect the outflowing jet, but also acts as proper test case for future simulations.
%}}

In the advective disc regime, numerical simulations \citep{mrc96,dcnm14,lckhr16}
and theoretical investigations \citep{cd07,kc13,kcm14,kc17,kscc13,ck16} showed that the extra-thermal gradient force in the post shock region automatically generates bipolar outflows.
Anticipating that the intense radiation from the accretion disc may accelerate jets, \cite{cc00a,cc00b,cc02a,cc02b}
investigated radiative driving of jets by advective disc photons.
It was noted that, cold jets could be efficiently accelerated to $\vt \sim$few$\times0.1$c.
But to achieve $\vt > 0.9$c for jets, the required base temperature and injection speed was quite high, which does not match with inner accretion disc parameters. Moreover, being in the non relativistic regime, the formalism followed by \citet{cc00a,cc02a} is only correct, up to the first order of the flow velocity.
In order to gauge the full extent of radiative acceleration,
investigations of radiatively driven particle jets in SR regime \citep{cdc04,c05}
were undertaken. Under such conditions, disc photons could accelerate jets up to $\gamt \gsim 2$ and
significant collimation could be achieved. The radiation field above such disc has two sources,
one from the inner post shock disc, which supplies the hard radiation, and two --- the soft radiation,
from the pre shock disc. It may be noted that, a compact, hot, low angular momentum corona close to the BH, which produces hard radiation and an external disc producing softer radiation, is not an exclusive of shocked advective discs but also of many other
models \citep{sl76,dwmb97,gzdjeuhp97}. Therefore, the source of radiation, i. e.,
the underlying accretion disc, may be an advective disc, or any other disc model which considers a compact, geometrically thick corona close
to the BH and an outer disc.

%{\textbf{\color{blue} Efforts have been made for study of jets 2-D 

In most of the investigations of relativistic fluid jets, the cross-section was assumed to be spherical ($\propto r^2$, $r$ being the radial distance). \cite{mstv04}, considered thermally driven relativistic jets
in Schwarzschild metric, modified an approximate equation of state (EoS) of single species relativistic gas \citep{m71}. They hid the actual acceleration process in an adhoc adiabatic index ($\Gamma$) and obtained monotonic jets from mildly to ultra relativistic jet terminal speed.
In contrast, \cite{ftrt85} studied jet driven by radiation, as well as, the cross-section deviated from spherical description. Since, the possibility of internal shocks in outflows, except
for non-spherical solar winds
\citep{lh90}, has not been reported very often, hence it was not clear whether non-conical geometry or
the external radiation field triggered the shock in the jet. 
Vyas et. al. (2015, hereafter VKMC15) addressed the problem of radiatively driven fluid jets in SR regime
similar to \cite{ftrt85}, but unlike it, used a relativistic EoS for the fluid and the jet geometry was conical. 
Although \cite{vkmc15} produced relativistic $\vt$, but no multiple sonic point or shock in jets were
obtained. 
We focussed on the role of jet geometry in Vyas \& Chattopadhyay (2017, hereafter VC17) and compared thermally driven relativistic jets with spherical cross-section with the non-spherical one. We showed, jets with non-spherical cross-section indeed produce
multiple sonic points and shock. However, there was no shock for flows with conical jets. 

In this paper, we revisit the problem as posed by \cite{ftrt85} and \cite{mstv04}, i. e.,
we consider radiatively driven jets like the former, but for conical jets like the latter,
such that no shock can form due to the flow geometry of the jet.
We use a relativistic EoS for a multispecies gas and solve the jet EoMs in curved geometry of
Schwarzschild metric. One of the main reason to use Schwarzschild metric instead of pseudo-Newtonian 
potential (pNp) in special relativistic metric, is because the curvature effect on the radiation field
is important and affects atleast up to few tens of gravitational radii
and also that the pNp makes the flow much hotter than real flows. Moreover, the radiative moments were re-computed from a thicker disc in the curved space-time, complete with all the transformations required to do so. It would be intriguing to study all possible jet solutions as the jet plies through the intense radiation field of the accretion disc.
Can radiation accelerate  jets to relativistic terminal speeds, starting with reasonable base temperature and speed, and whether accretion disc radiation can drive a jet shock. In this paper we would like to
investigate these questions.

In next section, we present the governing equations and underlying assumptions.
We also present the method to compute radiative moments from the approximate
accretion solutions and outline the solution methodology. Then we present the results (section \ref{sec3}). At the end, we conclude the paper discussing the outcomes and significance (section \ref{sec4}).

\section{Assumptions and governing equations}%\& structure of accretion disc}
\label{sec2}
\subsection{Assumptions}

The space-time around a non-rotating black hole is described by Schwarzschild metric:
\bea
ds^2=-g_{tt} c^2dt^2+
g_{rr}dr^2% \nonumber \\
+g_{\theta \theta}d{\theta}^2+g_{\phi \phi}d\phi^2  \nonumber \\
=-\left(1-\frac{2G\mbh}{c^2r} \right)c^2dt^2+
\left(1-\frac{2G\mbh}{c^2r}\right)^{-1}dr^2 \nonumber \\
+r^2d{\theta}^2+r^2\sin^2{\theta}d\phi^2
%\eqno{(1)}
\label{metric.eq}
\eea

Here $r$, $\theta$ and $\phi$ are usual spherical 
coordinates,
$t$ is time, $g_{\mu \mu}$ are diagonal metric components, $\mbh$ is the mass of the central 
black hole and $G$ is the universal constant of gravitation. Hereafter, we have used geometric units (unless specified otherwise) where $G=\mbh=c=1$ with, such that the units of mass, length and time are $\mbh$, $\rg=G\mbh/c^2$ and $\tg=G\mbh/c^3$, respectively. In this system of units, the event horizon or Schwarzschild radius is at $r_{\rm \small S}=2$. The jet is assumed to be in steady state ({\ie} $\partial/\partial t=0$). A jet cannot have high angular momentum, otherwise it will not remain collimated. Moreover, efficient removal of jet angular momentum by the radiation, has also been reported
before \citep{fth01,c05}, therefore, for simplicity we assume jets to be non-rotating ($u^{\phi}=0$), on-axis ({\ie} $u^\theta=0$) and axis-symmetric ($\partial/\partial \phi=0$) 
with small opening angle. Narrow jet allows us to further assume that at distance $r$, the physical variables of the jet remain same along the transverse direction. In this study, the jet is assumed to expand radially along the rotation axis of the accretion disc. 

The source of radiation is the accretion disc. The dominant radiative cooling processes considered
in the disc are synchrotron, bremsstrahlung and in addition, inverse-Comptonization
in the corona. The magnetic pressure in the accretion disc is assumed to
be due to stochastic magnetic field. The ratio of the gas pressure to the magnetic pressure is given by $\beta$.
We take $\beta=2.0$ in this paper. The cooling process in the corona is implemented through a fitting function
\citep{vkmc15}. This is an exploratory study of astrophysical fluid jets, which are powered by both the thermal gradient term and radiation driving. The accretion disc plays an auxiliary role, i. e., it influences the
jet only through radiation. The jet is assumed to be fully ionized and the interaction between radiation and matter is dominated by Thomson scattering. Full relativistic transformations are implemented on the radiation field. We use the methods laid down by  \cite{b02,bgjs15} to incorporate the effect of photon bending in computing radiative moments.
 %Effects of photon bending in radiation field are approximated with the help of} \cite{b02,bgjs15}, which accurately {\textbf{\color{blue} approximate} photon bending beyond $4-5\rg$.  

\subsection{Governing equations}

\subsubsection {Equation of state}
\label{sbsbsec2.1.1}
EoS is the relation between the thermodynamic quantities of fluid i. e., internal energy density ($e$), pressure ($p$) and mass
density ($\rho$). It is basically a
closure relation between the thermodynamic variables which allows us to solve
the equations of motion of a fluid.
In this study, we consider EoS for multispecies, relativistic
flow proposed by \citet{c08,cr09} which is an extremely close approximation
of the exact one \citep{c38,s57}. The EoS is 
given as,
\begin{equation}
e=n_{e^-}m_ec^2f, \mbox{ in physical dimensions}
\label{eos.eq}
\end{equation}
where, $n_{e^-}$ is the electron number density, $m_e$ is the electron rest mass and dimensionless quantity
$f$ is given by
\begin{equation}
f=(2-\xi)\left[1+\Theta\left(\frac{9\Theta+3}{3\Theta+2}\right)\right]
+\xi\left[\frac{1}{\eta}+\Theta\left(\frac{9\Theta+3/\eta}{3\Theta+2/\eta}
\right)\right].
\label{eos2.eq}
\end{equation}
Here,
$\Theta=kT/(m_ec^2)$ is a measure of temperature ($T$), $k$ is Boltzmann constant and
$\xi (= n_{p^{+}}/n_{e^{-}})$ being the relative proportion of number densities of 
protons and electrons.
$\eta (= m_{e}/ m_{p^{+}}$) is the mass ratio of electron and proton.
The expressions of the polytropic index $N$, adiabatic 
index $\Gamma$ and
adiabatic sound speed $a$ and enthalpy $h$ (in geometric units) are given by
\begin{equation}
N=\frac{1}{2}\frac{df}{d\Theta} ;~~ \Gamma=1+\frac{1}{N} ; ~~
a^2=\frac{\Gamma p}{e+p}=\frac{2 \Gamma \Theta}
{f+2\Theta}.; ~~ h=\frac{f+2 \Theta}{\tau}
\label{sound.eq}
\end{equation}
Here $\tau(=2-\xi+\xi/\eta)$ is a function of composition.

\subsubsection {Jet EoM}
Equations of motion i.e., EoM of radiation hydrodynamics in curved space-time, were derived before
\citep{p06,t07} and in the following, we present them in brief.
The energy-momentum tensor for the matter ($T^{\alpha \beta}_M$) and radiation ($T^{\alpha \beta}_R$)
are given by
\begin{equation}
T^{\alpha \beta}_M=(e+p)u^{\alpha}u^{\beta}+pg^{\alpha \beta};
~~T^{\alpha \beta}_R={\int}I_{\nu}l^{\alpha}l^{\beta}d{\nu}d{\Omega},
\end{equation}
here,  $u^{\alpha}$ are the components of four velocity
$l^{\alpha}$s are the directional derivatives, $I_{\nu}$ is the specific intensity
of the radiation field where $\nu$ is the frequency of the radiation and
$d\Omega$ is the differential solid angle subtended by a source point at the accretion disc surface on to the field point at the jet axis.

The $i^{\rm th}$ component of the momentum balance equation is obtained by projecting $(T^{\alpha \beta}_M+T^
{\alpha \beta}_R)_{;\beta}=0$ with the tensor
$(g^{i}_{\alpha}+u^iu_{\alpha})$
and in steady state it becomes
\begin{equation}
u^r\frac{du^r}{dr}+\frac{1}{r^2}=-\left(1-\frac{2}{r}+u^ru^r\right)
\frac{1}{e+p}\frac{dp}{dr}+{\rho}_e\frac{{\sigma}_{T}
}{m_e(e+p)}{\Im}^r,
\label{eu1con.eq}
%\eqno{(6a)}
\end{equation}

Here, $\rho_e$ is total lepton density and ${\Im}^r$ is the net radiative contribution\footnote{In physical units it is $\frac{\sigma_T}{m_ec}$ but in our unit system $c=1$} and is given by
\be
{\Im}^r=\sqrt{g^{rr}}\gamma^3\left[(1+v^2){\cal R}_1-v
\left(g^{rr} {\cal R}_0+\frac{{\cal R}_2}{g^{rr}}\right)\right]
\label{radcontrib.eq}
\ee
Three-velocity $v$ of the jet is defined as $v^2=-u_iu^i/u_tu^t=-u_ru^r/u_tu^t$, {\ie}
$u^r={\gamma}v\sqrt{g^{rr}}$ and $\gamma^2=-u_tu^t$ is the Lorentz factor. ${\cal R}_0,~{\cal R}_1$ and
${\cal R}_2$ are zeroth, first and second moments of specific intensity of the radiation and physically can be identified
as the radiation energy density, the flux and the pressure respectively.

In scattering regime, first law of thermodynamics, or energy equation ($u_{\alpha}T^{\alpha \beta}_{M_{;\beta}}=-u_{\alpha}T^{\alpha \beta}_{R_{;\beta}}$) is given by,
\begin{equation}
\frac{de}{dr}-\frac{e+p} {\rho}\frac{d\rho}{dr}=0,
%\eqno{(6b)}
\label{en1con.eq}
\end{equation} 
Therefore, the system is isentropic \citep{mm84}. 
Integrating the conservation of mass flux equation ($[\rho u^\alpha]_{;\alpha}=0$),
we obtain the mass outflow rate 
\begin {equation}
{\dot M}_{\rm o}=\rho u^r \cal A
\label{mdotout.eq}
\end {equation}
Here ${\cal A}(\propto r^2)$ is the cross-section of the jet.

Since r.h.s of the energy equation (\ref{en1con.eq}) is zero, then integrating
it with the help of the EoS (\ref{eos.eq}) we obtain an adiabatic relation between $\Theta$ and $\rho$
\citep{kscc13}. Replacing $\rho$ of the adiabatic relation into the equation (\ref{mdotout.eq}), we obtain 
expression of entropy-outflow rate
\begin{equation}
\mdtj=\mbox{exp}(k_3) \Theta^{3/2}(3\Theta+2)
^{k_1}
(3\Theta+2/\eta)^{k_2}u^rr^2,
\label{entacc.eq}
\end{equation}
where, $k_1=3(2-\xi)/4$, $k_2=3\xi/4$, and $k_3=(f-\tau)/(2\Theta)$.
This is also a measure of entropy of the jet and in the present context, it remains constant 
along the jet except at the shock.
We integrate equations (\ref{eu1con.eq} and \ref{en1con.eq}), and obtain the generalized relativistic Bernoulli
parameter in the radiation driven regime and is given by,
\be 
\begin{split}
& E=-h u_t{\rm exp}(-X_f),~~\mbox{where,} \\
& X_f=\left(\int dr \frac{\gamma (2-\xi)}{(f+2 \Theta )\sqrt{g^{rr}}}\left[(1+v^2){R_1}-v
(g^{rr} R_0+\frac{R_2}{g^{rr}})\right]\right).
\end{split}
\label{energy.eq}
\ee
Here, $R_0=\sigma_T{\cal R}_0/(m_e),~R_1=\sigma_T{\cal R}_1/(m_e)$ and $R_2=\sigma_T{\cal R}_2/(m_e)$
are terms proportional to the radiative moments like radiation energy density , flux and pressure, but for simplicity in rest of the paper, we call these quantities ($R_0,~R_1,~\&~R_2$) as respective radiative moments.
The kinetic power of a jet, is defined as the energy flux at large distances and is given as:

\be 
L_j=\dot{E}={\dot M}_{\rm o}{E_{\infty}}
\label{ljet.eq}
\ee
Here, $ E_\infty=[-hu_t]_{r\rightarrow \infty}$ is the Bernoulli parameter at infinity.

%\be 
%L_j=\dot{E}={\dot M}_{\rm o}(-hu_t)_{r\rightarrow \infty}
%\label{ljet.eq}
%\ee
Expressing $\wp^r=\sigma_T\Im^r/(m_e)$, equations (\ref{eu1con.eq}) and (\ref{en1con.eq}) can be expressed as gradients of $v$ and $\Theta$
and are given by
 \bea
\gamma^2vg^{rr}r^2\left(1-\frac{a^2}{v^2}\right)\frac{dv}{dr}=a^2\left(2r-3\right)-1+\frac{\wp^r r^2(2-\xi)}{(f+2 \Theta)\gamma^2}
\label{dvdr.eq}
\eea
and
\begin{equation}
\frac{d{\Theta}}{dr}=-\frac{{\Theta}}{N}\left[ \frac{{\gamma}
^2}{v}\left(\frac{dv}{dr}\right)+\frac{2r-3}{r(r-2)}
\right]
\label{dthdr.eq}
\end{equation}
Equations (\ref{dvdr.eq}) and (\ref{dthdr.eq}) are integrated to solve for $v$ and $\Theta$ of a steady jet
plying through the radiation field ($\Im^r$) of the underlying accretion disc.

The last term in the r.h.s of the equation (\ref{dvdr.eq}) is the radiation momentum deposition term,
\be
\frad=\frac{\wp^r r^2(2-\xi)}{\tau h \gamma^2}=\frac{\wp^r r^2(2-\xi)}{(f+2 \Theta)\gamma^2}
\label{radterm.eq}
\ee
with
$$
\wp^r=\sqrt{g^{rr}}\gamma^3\left[(1+v^2){R_1}-v
\left(g^{rr} R_0+\frac{R_2}{g^{rr}}\right)\right]
$$

%\be
%\frad=\frac{\wp^r r^2(2-\xi)}{(f+2 \Theta)\gamma^2}=\frac{r^2(2-\xi)}%{f+2\Theta}\sqrt{g^{rr}}\gamma\left[(1+v^2){R_1}-v
%\left(g^{rr} R_0+\frac{R_2}{g^{rr}}\right)\right]
%\label{radterm.eq}
%\ee

Equation (\ref{radterm.eq}) shows that, because of the presence of enthalpy in the denominator, the radiation driving of the jet is more effective for colder jets. The presence of the metric term $g^{rr}$ in
$R_d$ implies that gravity also affects radiation driving. One can reduce equation (\ref{radterm.eq})
to non-relativistic limits, if
$g^{rr} \rightarrow 1$, $\gamma^2 \rightarrow 1$ and $h \rightarrow 1$, then $\frad$ reduces to
\citep[also see,][]{cc02a,kcm14},
\be 
\frad {\rm (NR)}=\frac{r^2(2-\xi)}{\tau}[{R_1}-v(R_0+R_2)].
\ee
Clearly, $\frad ({\rm NR})$ is less interesting, since it is more dependent on the moments
and weakly on $v$.
Since the $g^{rr}$ term has appeared in $R_d$, therefore, closer to the horizon the third of the term
in equation(\ref{radterm.eq}) dominates. That is, as $r\rightarrow 2$,
$\frad \rightarrow -r^2v R_2 \gamma(2-\xi)/(h \tau \sqrt{g^{rr}})$, therefore,
the outward ($v>0$) moving jet will decelerate --- an effect that cannot be realised even with the special relativistic version of $\frad$ \citep{vkmc15}. An interesting comparison of equations of motion with Paczy\'nski-Wiita potential and general relativistic analysis is discussed in appendix B, where we show how pNp is insufficient for relativistic outflows and leads to deviation even at larger distances from BH. Impact of curved space on radiation field and radiative term is discussed separately in section (\ref{sec:rad_curved}). The resultant differences make general relativistic study inevitable for precise study of relativistic dynamics of jets.

Within the funnel for a geometrically thick corona, $R_1<0$ as will be shown later, and therefore,
within the funnel $\frad <0$ for outward moving jet i. e., $v>0$. But even in regions where $R_1>0$,
$\frad \leq 0$, for any $v\geq v_{\rm eq}$, where,
\be
v_{\rm eq}=\frac{(g^{rr}R_0+R_2/g^{rr})-\sqrt{(g^{rr}R_0+R_2/g^{rr})^2-4R^2_1}}{2R_1}
\label{equilbmv.eq}
\ee
It is clear from equation (\ref{equilbmv.eq}), the effect of radiation drag
is effective in optically thin medium (radiation penetrates the medium) and for distributed source.
The negative terms in $\frad$
depend on $v$ and hence is termed as a drag term.
One may compare the GR version of $v_{\rm eq}$ with the special relativistic and Newtonian versions
\citep{cc02a,cdc04}.

\subsection{Radiative moments}
\label{sbsec2.2}
\begin {figure}[h]
\begin{center}
 \includegraphics[width=9.cm]{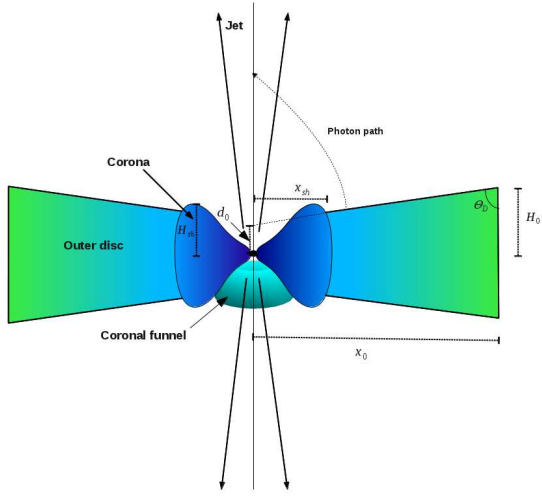}
\vskip -0.5cm
 \caption{Cartoon diagram of cross-sections of axis-symmetric
accretion disc and the associated jet in ($r,~\theta, \phi$ coordinates).
The shock location
$\xsh$, the intercept of outer disc on the jet axis ($d_0$), height of the shock $H_{\rm sh}$, the outer edge of the
disc $x_0$ are marked. Semi-vertical
angle of corona is $\theta_{\rm \small C}$ and for outer disc it is $\theta_{\rm \small D}$. The
funnel of the corona is also shown.}
%\vskip -0.75cm
\label{lab:fig1}
 \end{center}
\end{figure}

In Fig. \ref{lab:fig1},
we present the schematic diagram of the accretion disc-jet system, where the jet, the corona
and the outer disc are shown.
The outer boundary of the corona is $\xsh$ and the half height is $H_{\rm sh}$ and the outer boundary
of the outer disc is $x_0$ and the half height is $H_0$. As stated before, the accretion disc plays an auxiliary role in this paper, where it is considered only as a 
source of radiation. The accretion disc assumed, has a
geometrically thick, compact corona, which supplies the hard photons by inverse-Comptonization of seed photons,
and an outer disc supplying softer photons. Such a disc structure is broadly consistent with many
accretion disc models as has been mentioned in section \ref{sec1}. The Keplerian component in the outer disc
is ignored, because the radiative moments computed from an outer Keplerian
disc are negligibly small compared to those from the inner corona, or from the outer advective flow
\citep{cdc04,c05,vkmc15}. 
%\subsubsection{Estimating approximate accretion disc variables}

\subsubsection{Relativistic transformation of intensities from various disc components}
%\label{sbsbsec2.2.1}
To solve equations of motion of the jet, we need to compute radiative moments on the jet axis that requires information of specific intensities from both the outer disc and the corona. The details of estimating the temperature (\ref{acctemp.eq}) and velocity (\ref{accvel.eq}) from accretion discs and thereby estimating the radiative intensity (\ref{skint.eq}, \ref{coronaint.eq}), has been presented in appendix A. However, the form of the intensities is in the local rest-frame of the disc surface, and therefore,
those intensities need to be transformed from the disc rest frame 
to the curved frame. After special and general relativistic transformations the specific intensities
become, 
\begin{equation}
I_{\rm i}=\frac{{\tilde I}_{\rm i}}{\gamma^4_{\rm i}\left[1+{\vartheta}_jl^j\right]^4_{\rm i}}\left(1-\frac{2}{x}\right)^2
\label{Itrans.eq}
\end{equation}
Here ${\tilde I}_{\rm i}$ is the frequency integrated specific intensity measured in the local rest frame of the accretion disc, ${\vartheta}^j$ is $j^{\rm th}$ component of 3-velocity of accreting matter, $l^j$s are directional cosines, $\gamma_{\rm i}$ is Lorentz factor and $x$ is the radial coordinate of the source point on the accretion disc. 
The suffix ${\rm i}{\rightarrow {\rm \small C},~{\rm \small D}}$ signifies the
contribution from the corona and the outer disc, respectively. The presense of $(1-2/x)^2$ in the above equation
reduces the intensity of radiation close to the horizon \citep{b02}.

\subsubsection{Calculation of radiative moments in curved spacetime}
Radiative moments are defined as zeroth, first and second moments of specific intensity
i. e., $\int I d\Omega;~\int I l^j d\Omega;~\&~ \int I l^j l^k d\Omega$, respectively, which are
ten independent components \citep{mm84,c05}. However,
it was also found that for a conical narrow jet only three of the moments are dynamically important.  
%\textbf{\color{blue} Since} the space-time is curved around a black hole, therefore,
%the direction cosines need to be transformed in order to incorporate the curvature.
If $l_{\rm \small F}$ is the relevant direction cosine in the flat space-time, 
then it is related to the one in the curved space as \citep{b02}, 
\bea
l_{\rm i}=l_{\rm i \small F}\left(1-\frac{2}{x}\right)+\frac{2}{x} \nonumber \\
d \Omega_{\rm i} = \left(1-\frac{2}{x}\right) d \Omega_{\rm i \small F} 
\label{l_do_trans.eq}
\eea
Here, as before ${\rm i} \rightarrow {\rm \small C}~\&~ {\rm \small D}$ signifies disc components.

The expressions of flat space differential solid angle $d \Omega_{\rm i \small F}$ and direction cosines
$l_{\rm i \small F}$ are obtained to be
$$
d \Omega_{\rm i \small F}=\frac{rxd\phi dx}{[(r-x \cos\theta_{\rm i})^2+x^2\sin\theta_{\rm i}^2]^{3/2}}
$$

$$
l_{\rm i \small F}=\frac{(r-x \cos\theta_{\rm i})}{\sqrt{[(r-x \cos\theta_{\rm i})^2+x^2\sin\theta_{\rm i}^2]}}
$$
We use equations (\ref{Itrans.eq}) and (\ref{l_do_trans.eq}) in the definition of various radiative moments, and express
all the radiative moments ($R_0, ~ R_1~\&~R_2$) in a compact form given by,
\bea
R_{n\rm i}=\int^{x_{\rm i 0}}_{x_{\rm ii}} \int^{2\pi}_{0}\left(1-\frac{2}{x}\right)^3\frac{{\tilde I}_{\rm i}}{\gamma^4_{\rm i}\left[1+{\rm v}_jl^j\right]^4_{\rm i}} \nonumber \\
\times \left[\frac{(r-x \cos\theta_{\rm i})}{\sqrt{[(r-x \cos\theta_{\rm i})^2+x^2\sin\theta_{\rm i}^2]}}\left(1-\frac{2}{x}\right)+\frac{2}{x}\right]^n \nonumber \\
\times \frac{rxd\phi dx}{[(r-x \cos\theta_{\rm i})^2+x^2\sin\theta_{\rm i}^2]^{3/2}}
\label{moments2.eq}
\eea
Here limits of radial integration are $x_{\rm ii}$ (inner edge) and $x_{\rm i 0}$ (outer edge) of the respective disc component.
The index $n=0, 1, 2$ gives us
$R_0,~R_1~\&~R_2$, i. e., radiative energy density, radiative flux along $r$ and the $rr$ component
of the radiative pressure.
Since there are two disc components corona and outer disc, so at a given $r$ the net moments are,
\be 
R_n=R_{n{\rm \small C}}+R_{n{\rm \small D}}
\label{moments3.eq}
\ee

The $x$ limit of the corona are
$x_{\rm {\small C}i}=2, x_{{\rm \small C} 0}=\xsh$. However, from a given $r$, an observer cannot see the whole of the disc because the corona
blocks a portion of the disc. Therefore the inner edge of the outer disc is given by,
$$
x_{\rm {\small D}i}=\frac{r-d_0}{(r-H_{\rm sh})/\xsh+ \cot \theta_{\rm \small C}}
$$
It is clear from above that, as $r\rightarrow \infty$,
$x_{\rm {\small D}i}\rightarrow \xsh$. Moreover, up to some radius, radiation from the outer disc will never
reach the axis of the jet. If the distance above the disc up to which outer disc radiation does not reach the axis
is $r_{\rm lim}$, then
\be
r_{\rm lim}=\frac{x_0H_{\rm sh}-H_0\xsh}{x_0-\xsh}.
\label{shadolim.eq}
\ee

\subsection{Method of obtaining solutions}
\label{sec:method}
The jet solutions can be obtained by integrating equations (\ref{dvdr.eq} and \ref{dthdr.eq}).
Since, the jet originates from the accretion flow from a region close to the horizon,
therefore, the jet speed should be small but because of hot base, the jet base is subsonic.
At large distances from the BH, the jet moves with very high speed and is cold and hence it is supersonic.
So let the jet become transonic i.e, $v_c=a_c$ at the sonic point ($r=r_c$). Here suffix $c$ denotes quantities on the
sonic point.
Further, at $r_c$, $dv/dr\rightarrow 0/0$, which enables us to write down sonic point conditions as
\be 
v_c=a_c;\\ 
\label{sonic1.eq}
\ee
and 
\begin{equation}
a_c^2-\frac{1}{2r_c-3}+\frac{(\frad)_c}{2r_c-3}=0.
\label{sonic2.eq}
\end{equation}
At $r_c$, $dv/dr$ is obtained by L$^\prime$H\^opital's rule.
Equation (\ref{sonic2.eq}) gives functional dependence of the sound speed on $r_c$, from which $\Theta_c$ the temperature at the sonic
point can be easily obtained. $\Theta_c$ can be used to determine all other parameters at the sonic point like $a_c$, $\mdtjc$ (using equations \ref{sound.eq} and \ref{entacc.eq}). 
Since $E$ has no exact analytical form, it is obtained by numerical integration.
Moreover, $E$ is a constant of motion and $\mdtj$ an integration constant for the present case,
one can supply either and obtain the value of $r_c$, or, supply values of $r_c$ one may calculate all the flow
quantities, and start integrating using Runge–Kutta’s $4^{th}$ order method
from $r_c$, inwards and outwards to obtain the solutions.
To determine density, one may need to explicitly supply ${\dot M}_{\rm o}$ which are about
few percent of accretion rates, as has been theoretically obtained \citep{ck16,kc17}. 

%\subsubsection{Shock conditions}
%The existence of multiple 
%sonic points in the flow opens up the possibility of 
%formation of shocks in the flow. At the shock,
%the flow is discontinuous in density, pressure and velocity.
%The relativistic Rankine-Hugoniot conditions relate the flow quantities across the
%shock jump \citep{t48,cc11}
%\begin{equation}
%  [{\rho}u^r]=0,
%  \label{sk1.eq}
%\end{equation}
%\begin{equation}
%   [\dot{E}]=0
%   \label{sk2.eq}
%\end{equation}
%and
%\begin{equation}
%[T^{rr}]=[(e+p)u^ru^r+pg^{rr}]=0
%\label{sk3.eq}
%\end{equation}
%The square brackets denote the difference 
%of quantities across the shock, i.e. 
%$[Q]=Q_2-Q_1$ 
%with $Q_2$ and  $Q_1$ being 
%the quantities after and before the shock, respectively.\\
%
%Equation (\ref{sk2.eq}) states that the energy flux remains 
%conserved across the shock. Dividing 
%equation (\ref{sk3.eq}) and equation (\ref{sk2.eq}) by equation (\ref{sk1.eq}) and a little 
%algebra leads to
%\be
%\left[\left(h \gamma v+\frac{2 \Theta}{\tau \gamma v}\right)\right]=0;~\&~ [E]=0.
%\label{sk5.eq}
%\ee
%We check for shock conditions (equation \ref{sk5.eq}),
%as we solve the equations of motion
%of the jet.
\subsubsection{Shock conditions}
The existence of multiple 
sonic points in the flow opens up the possibility of 
formation of shocks in the flow. At the shock,
the flow is discontinuous in density, pressure and velocity.
The relativistic Rankine-Hugoniot conditions relate the flow quantities across the
shock jump \citep{t48,cc11}
\begin{equation}
  [{\rho}u^r]= [\dot{E}]=[T^{rr}_M+T^{rr}_R]=0,
  \label{sk1.eq}
\end{equation}
%The square brackets denote the difference 
%of quantities across the shock, i.e. 
%$[Q]=Q_2-Q_1$ 
%with $Q_2$ and  $Q_1$ being 
%the quantities after and before the shock, respectively.\\
%Equations (\ref{sk2.eq}) states that the energy flux remains 
%conserved across the shock.
Dividing ${\dot E}$ and $T^{rr}$ conservation conditions by mass conservation equation followed by a little 
algebra leads to
\be
\left[\left(h \gamma v+\frac{2 \Theta}{\tau \gamma v}\right)\right]=0;~\&~ [E]=0.
\label{sk5.eq}
\ee
We check for shock conditions (equation \ref{sk5.eq}),
as we solve the equations of motion
of the jet. However, one should note that unlike VC17, the thermal energy ($-hu_t$) doesn't remain conserved across the shock and the corresponding conserved quantity is generalized Bernoulli parameter $E$. 

\section{Analysis and results}
\label{sec3}
\subsection{Nature of radiative moments}
\begin {figure} %[h]
%\begin{center}
 \includegraphics[width=10.0 cm, trim=0 0 100 150,clip]{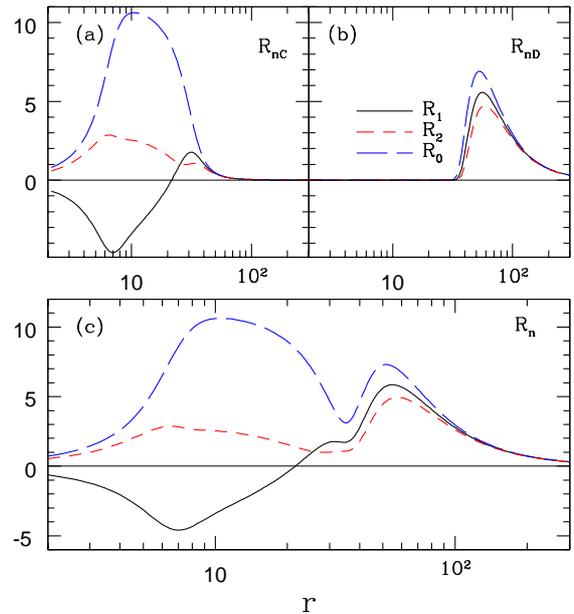}%
\vskip -0.2cm
 \caption{Distribution of radiative moments-energy density $R_0$ (long dashed, blue), flux $R_1$ (solid, black) and pressure
 $R_2$ (dashed, red) with $r$ above an accretion disc with $\dot{m}=10$. Radiative
 moments produced by various components of the disc, e. g., (a) from corona $R_{n \rm \small C}$; (b) from outer disc $R_{n \rm \small D}$ and (c) total radiative moments $R_n$}
\label{lab:fig2}
% \end{center}
\end{figure}
In Figs. (\ref{lab:fig2}a-c), we plot radiative energy density $R_0$
(long dashed, blue), flux $R_1$ (solid, black) and radiative pressure $R_2$ (dashed, red) as functions of $r$.
The components of the radiation field presented in all the panels are for $\dot{m}=10$ which corresponds to a size of corona
of $\xsh=12.31$ (see, equation \ref{xsdotm.eq}). The luminosity of such an accretion disc is $\ell=0.8$ around a BH of $\mbh=10 \msol$. In Figure (\ref{lab:fig2}a) we plot coronal moments $R_{n\rm \small C}$ (in compact notation) from discs around $\mbh=10 M_\odot$. 
The moments from the corona dominate the radiation field close to the BH.
And because the corona is geometrically thick, the radiation flux ($R_{1\rm \small C}$) is negative
within the funnel like region and therefore, is likely to oppose the jet flowing out, along with the
radiation drag terms (negative terms in r.h.s of equation \ref{radcontrib.eq}). Fig (\ref{lab:fig2}b) shows moments (presented in compact notation $R_{n\rm \small D}$) from the outer disc.
Because of the shadow effect from the corona, all moments of the outer disc are zero
for $r\leq r_{\rm lim}(=30)$
obtained from equation (\ref{shadolim.eq}). The moments of the outer disc for ${\dot m}$
peak around $r=55$.
In Fig. (\ref{lab:fig2}c), we plot the total radiative moments from the outer disc and the corona.
Far away from the BH ($r >$few$\times 10^2$), the jet sees the disc like a point source and all moments fall like inverse squared of the distance and at such distances $R_0 \sim R_1 \sim R_2$.
%In Figs. (\ref{lab:fig2}d), we compare total moments (in compact form $R_n$) computed in curved space time
%and in flat space-time. In presence of space-time curvature,
%a photon has to traverse greater distances than in a flat space-time, therefore leading to decrease in the %magnitude of the moments, as is shown in this panel. Moments with lower magnitude {\textbf{\color{blue} are} computed in curved space
%and have been used in this paper. 
\subsubsection{Effect of curved spacetime on radiation field and radiation drag}
\label{sec:rad_curved}
\begin {figure}[h]
\begin{center}
 \includegraphics[width=6.cm, trim=0 0 170 150,clip, angle=0]{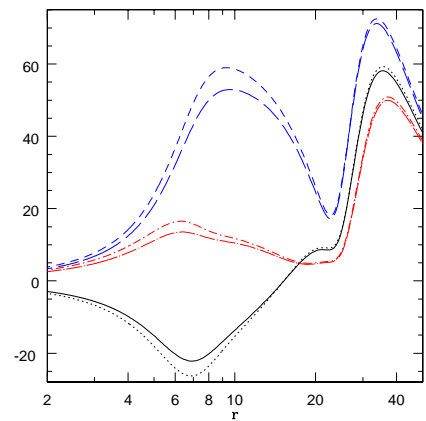}
\vskip -0.5cm
%\textbf{\color{blue}
 \caption{Energy density $R_0$ (long dashed, blue) $R_{0F}$ (dashed, blue), $R_1$ (solid, black) $R_{1F}$ (dotted, black), $R_2$ (long dashed-dotted, red) $R_{2F}$ (dashed-dotted, red) for $\ell=2.25$. Quantities with subscript $F$ denote moments calculated in flat space}%}
%\vskip -0.75cm
\label{lab:fig3}
 \end{center}
\end{figure}
\begin {figure}[h]
\begin{center}
 \includegraphics[width=8.cm, trim=0 0 100 150,clip, angle=0]{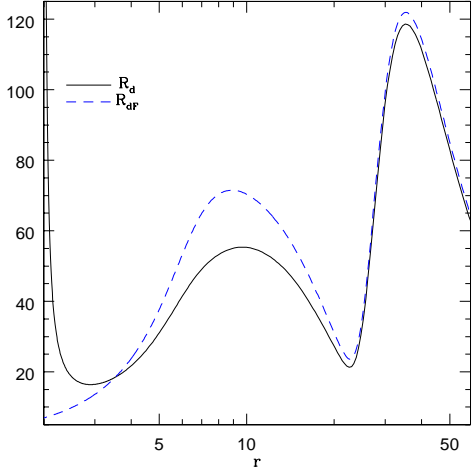}
\vskip -0.5cm
%\textbf{\color{blue}
 \caption{Comparison of radiation drag term $R_d$ computed in curved space (solid black) and flat space (dashed blue) for the moments shown in Fig. (\ref{lab:fig3})}.%}
%\vskip -0.75cm
\label{lab:fig4}
 \end{center}
\end{figure}
The radiation field in VKMC15 was calculated assuming flat space as pNp do not take care of impact of gravity in radiation fields. In Fig. (\ref{lab:fig3}), we compare radiative moments calculated in flat space with curved space for $\ell=2.25$. Various curves represent energy density $R_0$ (long dashed, blue), $R_{0F}$ (dashed, blue), $R_1$ (solid, black), $R_{1F}$ (dotted, black), $R_2$ (long dashed-dotted, red), $R_{2F}$ (dashed-dotted, red). Moments in curved space $R_{n}$ are different than that in flat space $R_{nF}$ because of the presence of metric components.
The metric components related to the accretion disc coordinates enter
inside integral while calculating radiative moments (equation \ref{moments2.eq}).
The appearance of $n$ as a power in equation (\ref{moments2.eq}) shows that the curvature effects are different for different moments. 
Further, the metric component $g^{rr}$ appears inside the radiative term while determining $\wp$. So the curvature affects the radiative term
in a very complicated way. In order to quantify the difference curvature has on the radiative terms, we compare the
radiation drag term $R_d=g^{rr} {\cal R}_0+\frac{{\cal R}_2}{g^{rr}}$ (solid, black) in the curved space with its version in the flat space $R_{dF}={\cal R}_{0F}+{\cal R}_{2F}$ (dashed, blue) in Fig. (\ref{lab:fig4}), for the same luminosity as in Fig. (\ref{lab:fig3}).
%($n=0,1,2$ represent ${\cal R}_{0}, {\cal R}_{1}, {\cal R}_{2},$ respectively). As we explained in paper that one contribution comes for transformation of specific intensity (equation 27) which cuts down the radiative contribution, while others appear for transformation of solid angle and direction cosines (equation 28). The cubic term of redshift factor alone says that even at $r=25$ the moments are cut down by 22$\%$. 
The difference is clearly visible. At $r\gsim 2$ the drag term $|R_d|>>|R_{dF}|$, but at $r>3.5$ the curvature
effect changes in an opposite manner i. e., $R_d < R_{dF}$. At $8\sim r \sim 9$, $R_{dF} \sim 1.3R_d$ which is the maximum deviation from the curved space values. However, the most interesting thing is that the drag term computed in the flat space is about three percent more than that computed in the curved space, even at a distance of about hundred gravitational radii. In other words, not only the curvature affects the radiative moments at moderately large distance, but since deviation varies with distance, one cannot use a scale factor to co-opt the curvature effect on radiation in flat space.
\subsection{Nature of sonic points}
\label{sec:sonic_points}
\begin {figure} %[h]
%\begin{center}
\vskip -1.5cm
 \includegraphics[width=12.cm, trim=0 0 0 80,clip]{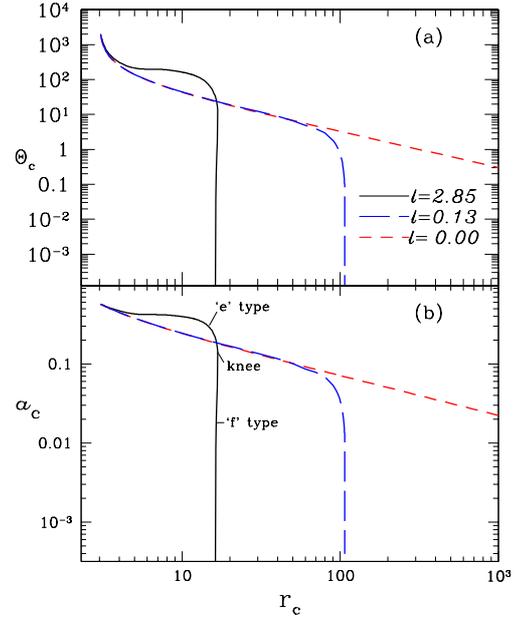}%trim=190 0 0 0,clip
\vskip -0.5cm
\caption{Variation of (a) $\Theta_c$ and (b) $a_c$ with $r_c$ for a jet acted on by $\ell=2.85$
(solid, black), $0.13$ (long dashed, blue) and thermal jet (dashed, red). The jet is composed of electrons and protons ($\xi=1$).}
\label{lab:fig5n}
% \end{center}
\end{figure}
\begin {figure}%[h]
%\begin{center}
 \includegraphics[width=12.cm, trim=0 0 0 200,clip]{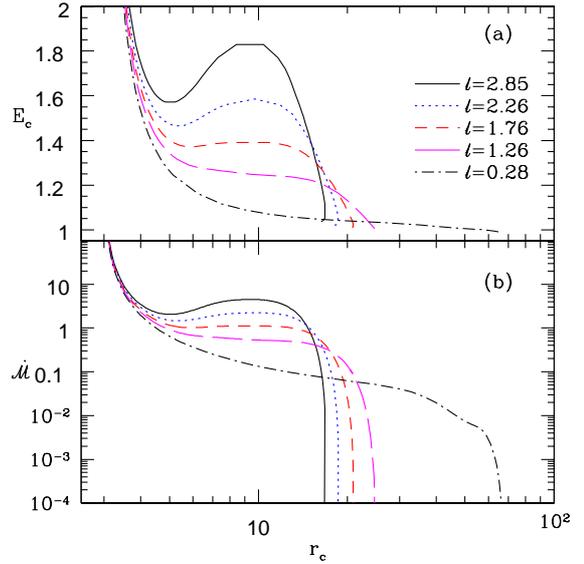}%trim=190 0 0 0,clip
\vskip -0.5cm
\caption{(a) $E_c$ and (b) ${\dot {\cal M}_c}$ as functions of $r_c$. Various curves represent $\ell=2.85$ (solid, black), $\ell=2.26$ (dotted, blue), $\ell=1.76$ (dashed, red), $\ell=1.26$ (long-dashed, magenta) and $\ell=0.28$ (dash-dotted, black).} 
%\vskip -0.75cm
\label{lab:fig6n}
% \end{center}
\end{figure}
We present $\Theta_c$ (Fig. \ref{lab:fig5n}a) and $a_c$ (Fig. \ref{lab:fig5n}b) as functions of $r_c$. 
Each plot represents sonic point properties of jets in a radiation field of an accretion disc
with luminosities $\ell=2.85$ (solid, black), $\ell=0.13$ (long dashed, blue) and $\ell=0.0$ or thermally driven jet
(dashed, red). Physically, different values of $r_c$ imply different choices of boundary conditions that give different transonic solutions. 
In absence of radiation, equation (\ref{sonic2.eq}) reduces to sonic point condition for thermal jets [$a_c^2=1/(2r_c-3)$]. This implies, for the physical values of $a_c$ i. e., $1/{\sqrt3}>a_c>0$, the range of
sonic point is $3\rg<r_c<\infty$.  
In the presence of radiation, the range of sonic point reduces to $3<r_c<$ some finite distance, as shown in
Figs. (\ref{lab:fig5n}a, b). The case with $\ell=0.13$ (long dashed, blue) almost follows the
curve for thermal jets (dashed, red) till about $50\rg$ but then it deviates and terminates at a distance
$\sim 100\rg$. The sonic point properties (i. e., $\Theta_c$ and $a_c$) for $\ell=2.85$ (solid, black) are significantly different from the thermal jet (dashed, red) and terminate at $14\rg$.

%For $\ell=2.85$ the curve deviates from the thermal one in few $\rg$ and terminates 
It is worth mentioning that in VKMC15, there were no sonic points between $3$---$4\rg$. Hence solutions in the present paper in which sonic
points are in the range $3\rg<r_c<4\rg$, cannot be found in VKMC15 (Appendix B.2).
This is because using pNp to mimic strong gravity makes the flow unphysically hot.
As a result there is enhanced thermal acceleration in all the solutions of VKMC15 compared to the
present one. This highlights one of the drawbacks of gluing special relativistic analysis with Paczy\'nski-Wiita potential. 

The $a_c-r_c$ curve in Fig. (\ref{lab:fig5n}b) form a `knee' like structure and rapidly decreases such that
at some $r_c\rightarrow r_{c\rm f}$, $a_c \rightarrow 0$. At the `knee' $da_c/dr_c \rightarrow \infty$ and the curve
bulges slightly, although not perceptible in the figure. 
Truncation of $r_c$ was also seen in special relativistic \citep{vkmc15} and pseudo-Newtonian studies \citep{cc00a} of radiatively driven jets. The estimation of $r_{c\rm f}$ can be obtained from equations
(\ref{sonic1.eq}, \ref{sonic2.eq}) by imposing $a_c \sim$small,
\be
\frac{(2-\xi)r_{c\rm f}^{3/2}(r_{c \rm f}-2)}{\tau}R_{1c\rm f}=1.
\label{sonic3.f}
\ee
In this paper, all the solutions corresponding to the sonic points under the `knee' are called `f'-type solutions while solutions above `knee' are referred to as `e'-type solutions, as marked in Fig. (\ref{lab:fig5n}b).

In Fig. (\ref{lab:fig6n} a-b) we plot $E_c$ and $\mdtjc$ as functions of $r_c$ respectively. Various curves correspond to $\ell=2.85$ (solid, black), $\ell=2.26$ (dotted, blue), $\ell=1.76$ (dashed red), $\ell=1.26$ (long dashed magenta) and $\ell=0.28$ (dashed dotted black). \cite{vc17} showed that for thermal flows with conical jet geometry, $E_c$ and $\mdtjc$ were found to be monotonic functions of $r_c$. In this paper, Figs. (\ref{lab:fig6n} a-b) show that $E_c$ and $\mdtjc$ of radiatively driven conical jets are non-monotonic functions of $r_c$. Above a certain value of $\ell$ (Fig. \ref{lab:fig6n} a), each curve has a maximum and a minimum. For a given
$E=E_c$ and $\ell$ within the two extrema, there is a possibility of forming three sonic points
(for curves with parameters $\ell=2.85,~2.26,~1.76$), where inner and outer sonic points are saddle-type, while middle sonic points are of spiral type. %The spiral nature of the sonic point is evident because of the thermal energy parameter ($-hu_t$) is not conserved along the flow.
Each of the sonic points for a given $E~\&~\ell$ have different entropy ($\mdtjc$).
Similarly, for a given choice of $\mdtj=\mdtjc$ and $\ell$ (Fig. \ref{lab:fig6n} b),
there is a possibility of three sonic points, differentiated by $E_c$.

\subsection{Jet solutions}
\begin {figure}[h]
\begin{center}
 \includegraphics[width=9.5cm, trim=0 0 0 110,clip]{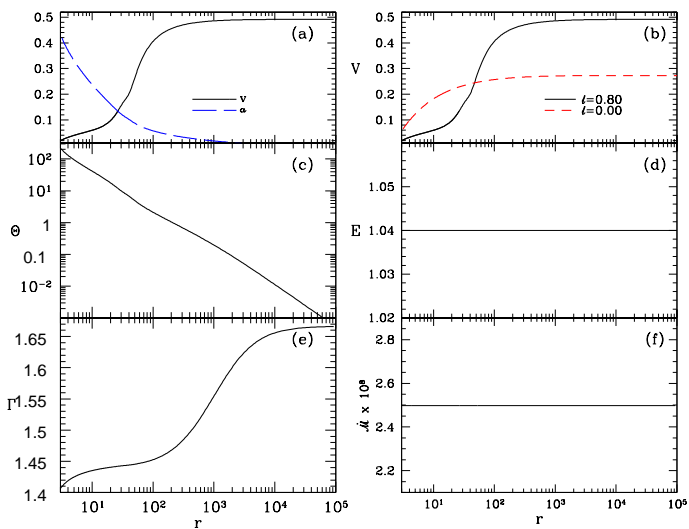}%
%\vskip 0.5cm
 \caption{(a) Three velocity $v$ (solid, black) and sound speed $a$ (long dashed, blue) as functions of $r$.
(b) Comparison of velocity distribution of thermally driven or $\ell=0$ jet
(dashed, red) and the radiatively driven jet (solid, black);
(c) $\Theta$; (d) $E$; (e) $\Gamma$ and (f) ${\dot {\cal M}}$
as a function of $r$. All the plots are for $E=1.04$ and radiatively driven jet
is for $\ell=0.8$.  } 
%\vskip -0.75cm
\label{lab:fig7n}
 \end{center}
\end{figure}

\begin {figure}%[h]
\begin{center}
 \includegraphics[width=10.cm, trim=50 0 0 0,clip]{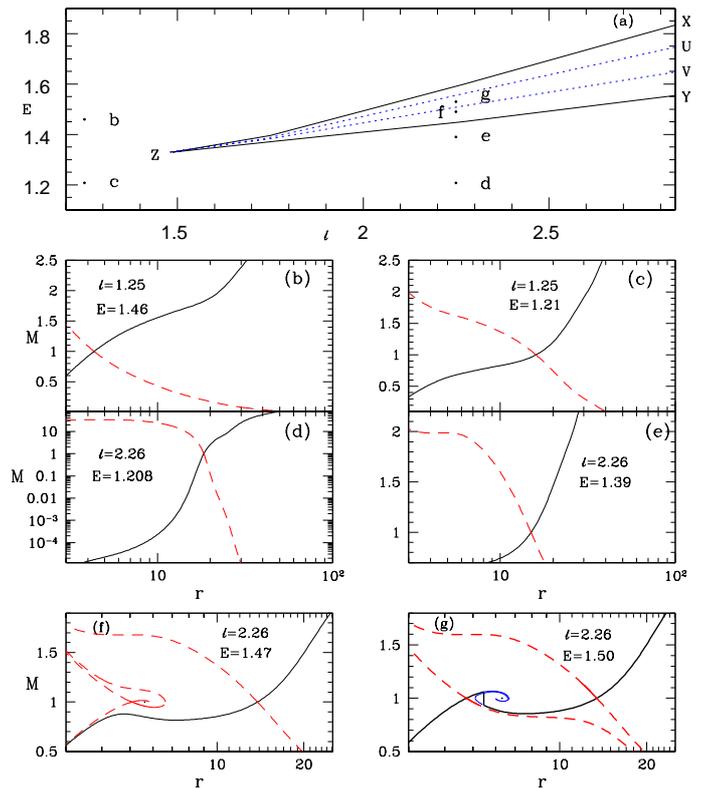}%
\vskip 0.5cm
 \caption{(a) $E-\ell$ parameter space: bounded region XZY signifies parameters for multiple sonic points in jet and region UZV within blue dotted lines represents parameters for which flow goes through shock transition.
 Filled circles named as `b-g' are the flow parameters $E$ and $\ell$, for which the jet solutions
 are plotted in panels (b)-(g). Mach number $M=v/a$ is plotted as a function of $r$ for (b) $E=1.46,~\ell=1.25$
 or the point b in panel a; (c) $E=1.208,~\ell=1.25$ or, point c in panel a; (d) $E=1.208,~\ell=2.26$
 or, point d in panel a; (e) $E=1.39,~\ell=2.26$ or point e in panel a; (f) $E=1.47,~\ell=2.26$ or, point f
 in panel a; and (g) $E=1.5,~\ell=2.26$ or point g in panel a.
 Each panel shows physical jet solutions (solid, black) and
 corresponding inflow solutions (dashed, red). Sonic points are shown by the crossing of inflow and jet
 solutions. All solutions are for $\ep$ flow.} 
\label{lab:fig8n}
 \end{center}
\end{figure}

We follow procedures of section \ref{sec:method} to obtain jet solutions and in Figs. (\ref{lab:fig7n}a-d) we present 
a typical jet solution characterized by generalized Bernoulli parameter $E=1.04$ and the composition of the
flow is $\xi=1$ or $\ep$ flow. In Fig. (\ref{lab:fig7n}a), three velocity $v$ (solid, black) and sound speed
$a$ (long dashed, blue) are plotted. The jet is transonic, starting with low $v$ and high $a$ and ending with the opposite. Interestingly, $R_1>0$ for $r>20$ above a disc and the jet starts
to accelerate significantly above that distance. The radiation field is for
$\ell=0.8$. 
In Fig. (\ref{lab:fig7n}b), we compare $v$ of a thermally driven jet (dashed, red) and radiatively driven
jet (solid, black), where $\vt$ of radiatively driven jet is about twice more than that the thermal jet.
The temperature of the radiatively driven jet decreases by five orders
of magnitude over a distance scale of five orders of $\rg$ (Fig. \ref{lab:fig7n}c) and consequently $\Gamma$
increases from a relativistic value to a non-relativistic one (Fig. \ref{lab:fig7n}e). The constant of motion
$E$ is plotted in Fig. (\ref{lab:fig7n}d) and since the flow is isentropic, ${\dot {\cal M}}$ is also constant (Fig. \ref{lab:fig7n}f). 

%The $E_c$---$r_c$ relation (Fig. \ref{lab:fig6n}a) show that for a luminous disc,
%the radiation may resist the flow at some regions, but drive the flow in some other.
Radiation from a luminous disc resists the jet within some distance above the funnel of the corona, but drives the flow beyond it. 
As a result, multiple sonic points are formed in jets at high $\ell$ for all $E$ within the maxima and minima
in each of the $E_c$---$r_c$ curves (Fig. \ref{lab:fig6n}a). Therefore, the loci of the maxima and the minima marks the range of
$E$ and $\ell$
for which the flow harbours multiple sonic points demarcated by XYZ in Fig. (\ref{lab:fig8n}a).
The region UZV (dotted, blue) represents flow parameters for which a jet has stable shock solution.
In Figs. (\ref{lab:fig8n}b-g) we plot the Mach number $M=v/a$ as a function of $r$,
where each panel corresponds to the coordinate points marked as `b'---`g' in $E$---$\ell$ parameter space
in Fig. (\ref{lab:fig8n}a). Here all possible jet solutions are presented (solid, black), but for the sake of completeness, we have also plotted the inflow solutions (dashed, red). The crossing points denote the locations
of the X-type sonic points. If the jet is illuminated by low luminosity radiation, then it flows out through
only one sonic point (Figs. \ref{lab:fig8n}b, c). If the jet is driven by high luminosity radiation,
then for lower energies, it will pass through a single outer type sonic point (Figs. \ref{lab:fig8n}d, e).
But for higher $\ell$ and $E$, the jet may posses multiple sonic points (Fig. (\ref{lab:fig8n}g, f).
In Fig. (\ref{lab:fig8n}g) the jet undergoes shock transition, but in Fig. (\ref{lab:fig8n}f)
it flows out only through the outer sonic point. The inner and outer sonic points are X type
and the middle one is spiral type (Figs. \ref{lab:fig8n}f \& g).
Figure (\ref{lab:fig8n}d) is of special importance,
since these are `f' type jets which start with very low velocities but achieve relativistic terminal speeds.

\begin {figure}[h]
\begin{center}
 \includegraphics[width=9.5cm, trim=0 5 0 150,clip]{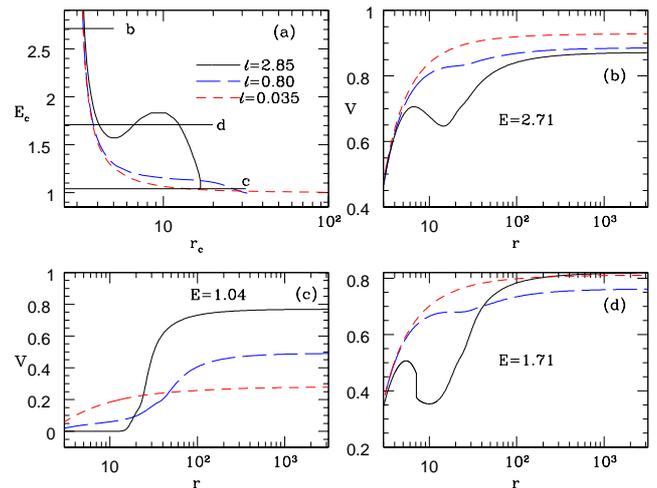}%
\vskip -0.5cm
 \caption{(a) $E_c$---$r_c$ plot with energy levels marked as `b' at $E=2.71$, `c' at $E=1.04$ and 
 `d' at $E=1.7$.
 (b) Comparison of three-velocity $v$ as a function of $r$ of jets starting with $E=2.71$.
 (c) Comparison of $v$ for `f'-type jets with $E=1.04$;
 and,
 (d)  Comparison of $v$ for jets with $E=1.7$.
 Each curve curve corresponds to $\ell=2.85$ (solid, black), $0.8$ (long dashed, blue) and
 $0.035$ (dashed, red).
 The composition of the jet is $\xi=1$ or $\ep$.} 
%\vskip -0.75cm
\label{lab:fig9n}
 \end{center}
\end{figure}

It is interesting to note that, the radiation effect is more perceptible for low energy jets than the
higher energy ones. To elaborate,
we once again invoke the $E_c$---$r_c$ curve in Fig. (\ref{lab:fig9n}a) for jets acted on by three disc luminosities
$\ell=2.85$ (solid, black), $\ell=0.8$  (long dashed, blue),
 $0.035$ (dashed, red), and mark three energy values as `b' at $E=2.71$, 'c' at $E=1.04$, and
 `d' at $E=1.7$. We compare the jet solutions at each of these values of $E$ in
 panels b, c and d of Fig. (\ref{lab:fig9n}).
 At high energies (i. e., Fig. \ref{lab:fig9n}b), radiation has no driving power due to presence of enthalpy
 in the denominator of the radiation term (equation \ref{radterm.eq}). The thermal gradient term
 in such cases is so strong that it accelerates the jet close to its local $v_{\rm eq}$ (equation \ref{equilbmv.eq}).
Therefore, shining radiation will only increase the radiation drag term and reduce the speed, as is seen
in this panel. Near the base, jets for all three $\ell$ achieve almost same $v$. As the temperature falls
and $\frad$ starts to become effective, jets plying through higher radiation field are slower (long dashed and solid curves).
%In  Fig. (\ref{lab:fig9n}c), we plot jets driven by radiation fields of $\ell=2.85$ (solid, black),
%$0.8$ (long dashed, blue) and $0.035$ (dashed, red), but all the jets have same $E=1.04$.
Radiation is
quite effective for low energy jets (Fig. \ref{lab:fig9n}c). Within the funnel $R_1$ is negative, therefore, the more is the disc
luminosity, greater will be the deceleration of jets inside the funnel. 
But above the funnel where $R_1>0$, radiation from luminous disc will drive
jets to higher terminal speeds. 
For middle energies e. g., $E=1.71$ (Fig. \ref{lab:fig9n}d), the effect of radiation is even more intriguing.
In presence of low luminosity radiation field, jets with moderate energies are thermally driven to achieve relativistic terminal speeds which are similar to the value achieved by purely thermally driven jet. Increasing
$\ell$, increases radiation drag and the jet speeds are suppressed, reducing the terminal speed. But for even higher $\ell$, the negative $R_1$ is strong enough to cause a shock transition
in the jet. In the post shock flow, because $v$ is significantly less than $v_{\rm eq}$, therefore, there is significant acceleration and roughly achieves the terminal speed of the thermally driven jet.
Therefore, for fluid jet, the role of radiation momentum deposition has multiple consequences
with distinctly different outcome, which underlines the importance of this study.

\begin {figure}[h]
\begin{center}
 \includegraphics[width=9cm, trim=0 0 160 10,clip]{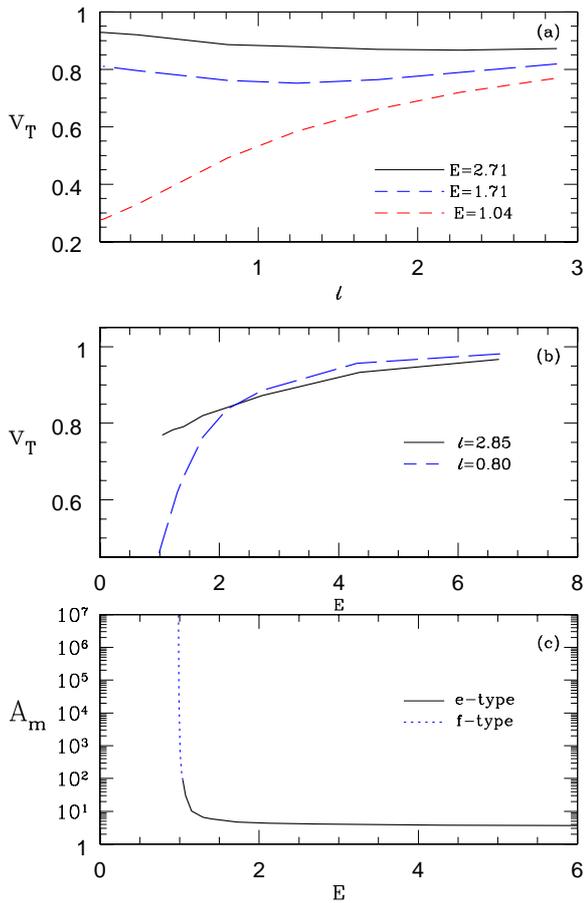}%
\vskip -0.5cm
\caption{(a) Variation of $\vt$ with $\ell$. Various curves represent $E=2.71$ (solid, black)
$E=1.71$ (long dashed, blue) and $E=1.04$ (dashed, red).
(b) $\vt$ as a function of $E$ for different $\ell=2.85$ (solid black) and $0.8$ (dashed, red).
(c) Amplification factor A$_{\rm m}$ with $E$ of jets flowing out through a radiation field of $\ell=0.8$. All panels have $\xi=1$. The `e'-type (solid) and `f'-type (dotted) jets are marked too.}
%\vskip -0.75cm
\label{lab:fig10n}
 \end{center}
\end{figure}

The definition of terminal speed or $\vt$ is the asymptotic jet speed, i. e.,
at $r\rightarrow$large, $v \rightarrow \vt$ where $dv/dr\rightarrow 0$. In Fig. (\ref{lab:fig10n}a),
we plot $\vt$ of jets with $\ell$ for three energies $E=2.71$ (solid, black), $E=1.71$ (long dashed, blue)
and $E=1.04$ (dashed, red). For low energy jets, terminal speed increases with $\ell$ (dashed, red). While for very high energy jets,
radiation drag decelerates the jet and $\vt$ decreases with $\ell$ (solid, black).
For moderate values of $E$, radiation
decelerates the jet when $\ell$ is low, but for higher $\ell$, $R_1$ within the funnel opposes the
outflowing jet to such an extent, that it triggers a shock transition. In the post-shock jet, $v$ is significantly less than $\veq$ and $R_1>0$, therefore radiation accelerates the jet efficiently to achieve high $\vt$.
In Fig. (\ref{lab:fig10n}b), we plot $\vt$ as a function of $E$, where each curve represents $\ell=2.85$ (solid, black) and $\ell=0.8$ (long dashed, blue). 
Similar to the previous panel, we find $\vt$ increases with $\ell$ for lower $E$ and
decreases for higher $E$.
It is interesting that for high $E$, $\vt$ is greater for lower $\ell$.
We also define an amplification parameter A$_{\rm m}=\vt/v_b$ as a measure of acceleration of the jet,
where $v_b$ is the base speed with which the jet is launched. 
In Fig. (\ref{lab:fig10n}c), we plot A$_{\rm m}$ as a function of $E$ for $\ell=0.8$. The dotted part of the
curve represents `f'-type solutions and the solid curve represents `e'-type solutions. It is clear from
the plot of the amplification parameter that,
radiation driving is more effective for `f'-type solutions, compared to the `e'-type jets.

\begin {figure}[h]
\begin{center}
 \includegraphics[width=8.cm, trim=0 0 220 10,clip]{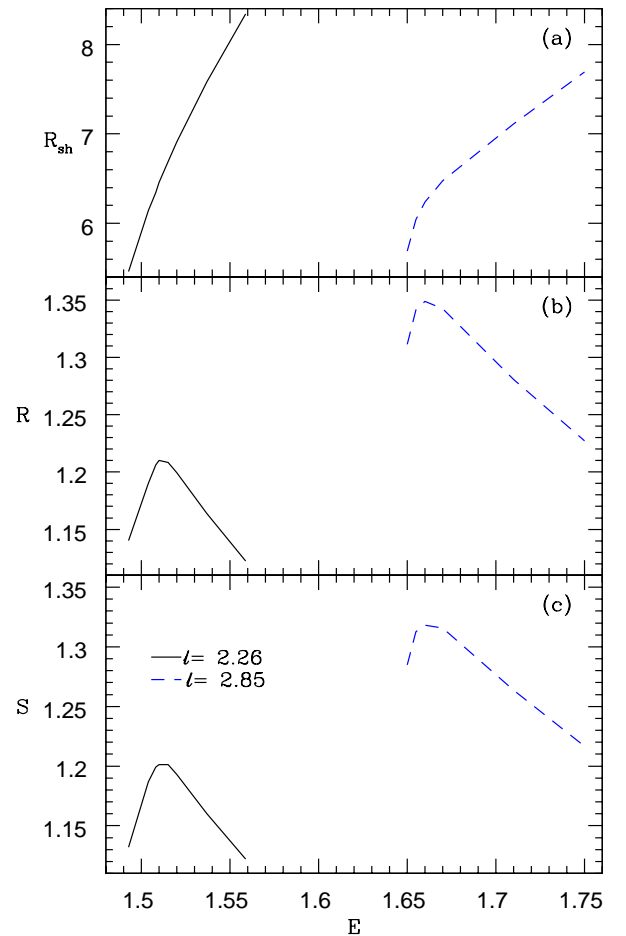}%
\vskip -0.5cm
 \caption{(a) Jet shock location $R_{\rm sh}$; (b) Compression ratio $R$
 and (c) shock strength $S$ as a function of $E$ for jets with composition $\xi=1.0$.
 Each curve is for $\ell=2.26$ (solid) and $\ell=2.85$ (long-dashed).} 
\vskip -0.75cm
\label{lab:fig11n}
 \end{center}
\end{figure}

Since the jet also contains radiation driven shock, so we plot the shock location $R_{\rm sh}$ (Fig. \ref{lab:fig11n}a), compression ratio $R$ (Fig. \ref{lab:fig11n}b), and shock strength $S$
(Fig. \ref{lab:fig11n}c) as a function of$E$ with each curve plotted for constant values of
$\ell$.
The compression ratio is defined as $R=\rho_+/\rho_-$ --- ratio of post and pre-shock mass densities; and the shock strength $S=M_-/M_+$ --- the ratio of pre and post-shock Mach numbers.
The composition of the jet is $\xi=1.0$ and each curve
is for $\ell=2.26$ (solid) and $\ell=2.85$ (long-dashed). 
In general, $R_{\rm sh}$ increases with $E$, because higher $E$ implies higher thermal energy at the base which pushes the shock front outwards. In jets, as the shock moves outwards the jump condition becomes steeper
and hence the shock becomes stronger. \cite{vc17}, which also showed the existence of shocks, was consistent with the above fact. However, the crucial difference between \cite{vc17} and the present venture is the agency that drive the shock. In \cite{vc17}, the shock is driven by the geometry of the flow and is coupled with the thermal term \citep[the coefficient of $a^2$ in equation 17 of][]{vc17} and therefore, the shock becomes stronger with
$E$. In the present paper, the shock is driven by the radiation that opposes
the jet flow within the funnel of the disc. In addition, the radiation term $\frad$ is more effective for flows with
lower thermal content i.e., with lower $E$. Therefore, increasing $E$ would negate the effectiveness of radiation,
and should weaken the shock. So $R$ and $S$ which measure shock strength, initially increase but eventually decrease with increasing $E$, maximizing at some value of $E$ in stark contrast with \cite{vc17}.
It is also quite clear that for higher $\ell$, the shock generally becomes stronger (long-dashed and solid curves).

\begin {figure}[h]
\begin{center}
 \includegraphics[width=9.5cm, trim=10 0 0 250,clip]{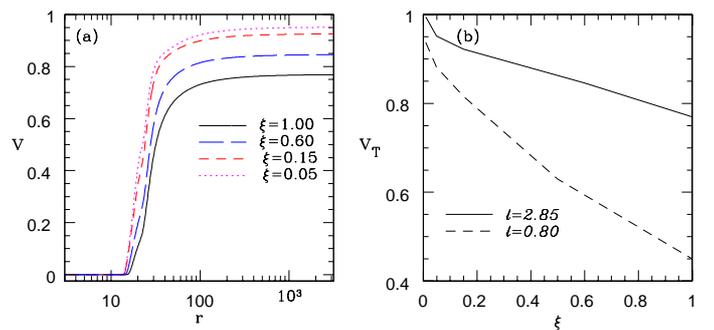}%
\vskip -0.5cm
 \caption{(a) Plot of $v$ as a function of $r$ for jets acted on by disc radiation marked by $\xi=1.0$ (solid, black); $\xi=0.6$ (long-dashed, blue), $\xi=0.15$ (dashed, red) and $\xi=0.05$ (dotted, magenta).
 The disc radiation is for $\ell=2.85$. (b) Variation of $\vt$ with $\xi$ for `f'-type jets, driven by radiation
 quantified by $\ell=2.85$ and $\ell=0.8$.} 
\vskip -0.75cm
\label{lab:fig12n}
 \end{center}
\end{figure}

A closer look into equation (\ref{radterm.eq}), reveals that $\frad$ is twice as large
for $\el$ jets than for $\ep$ jets for the same values of $\Theta$ and $v$.
Earlier it has been shown than lepton dominated flows are colder that $\ep$ flows \citep{cr09,cc11},
which means that the term $f+2\Theta$ is lower for low $\xi$ flow.
In other words, $\frad$
will be more effective for lepton dominated jets. However, one cannot compare jets with same $E$ across a
range composition. If one considers equation (\ref{energy.eq}), then one can easily understand that, a slight change in $X_f$
will affect the value of $E$ by a large amount. Since for low $\xi$ flow, $\Theta$s are quite different
than those of $\ep$ flow, therefore, jets with different $\xi$, starting with similar temperature and velocity,
will have widely differing $E$.
In Fig. (\ref{lab:fig12n}a) we compare jets launched with the same velocity at the base and driven
by radiation of same luminosity ($\ell=2.85$), each curve corresponds to $\xi=1.0$ (solid, black), $\xi=0.6$
(long-dashed, blue), $\xi=0.15$ (dashed, red) and $\xi=0.05$ (dotted, magenta). Jet speeds are higher for flow with lower $\xi$. In Fig. (\ref{lab:fig12n}b), we plot $\vt$ of the jet with the flow composition $\xi$,
each curve corresponds to super-Eddington luminosity (solid, $\ell=2.85$) and sub-Eddington luminosity
(dashed, $\ell=0.8$). For lepton dominated flow, the terminal speed can easily go above 90\% the speed of light.
%{\textbf{\color{blue}
%\subsection{On the nature of shocks}
%The shocks presented in this paper are manifestation of radiation field. The supersonic flow is resisted by %negative flux from the corona and induces shock transition in the jet. Jets in VC17 were thermal in nature. The shocks in VC17 were manifestation jet geometry where pinch off effect in supersonic branch enhances pressure in supersonic region and jet goes through shock transition. VC17 concluded that in absence of radiation, radial jets cannot form shocks. While here we show that even radial jets go through shock transition under the impact of radiation. It is to be noted that these shocks are different in nature as the reported shocks in the jets are either generated due to mechanical interaction between relatively moving plasma blobs or because of interaction of jets with ambient medium.
%The radiation driven shocks obtained in this paper are weaker compared to VC17. 
%}}
\section{Discussion and Conclusions}
\label{sec4}

In this paper, we have studied radiatively and thermally driven jets with spherical cross section having a small opening angle around BH. Since the flow is hot enough to be fully ionized, the momentum transferred from radiation to the jet is only
through scattering. The thermodynamics of the jet is described by a relativistic EoS, while it flows through the radiation field of the accretion disc
in Schwarzschild metric. 
The disc assumed, has a thick compact corona, which emits through bremsstrahlung and synchrotron processes
like the outer disc, but additionally, through the inverse-Compton process, all of which is implemented via 
a fitting function. 

Generally, most of the studies on radiatively driven jets are conducted in SR regime
and stronger gravity is mimicked by adding any gravitational potential adhoc in the momentum balance equation
\citep{ftrt85,vkmc15}.
Even if we over look the obvious mistake of gluing SR and any gravitational potential from the view point
of the famous Principle of Equivalence, still it produces many unphysical phenomena in the solutions.
For example the adhoc gravity in SR regime jet solutions become unrealistically hot, such that sonic points do not form within four Schwarzschild radii. Even in cases where transonic solutions are obtained, the thermal gradient term dominates completely the radiation term. This accelerates the jets to reach their local equilibrium velocity. Hence further out,
when the jet is cooler, radiation drag becomes more important than radiation driving. In proper GR regime, the radiation drag at moderate distances is much lower.

Since
we are considering curved space-time in the present paper, consequently the radiative moments
have been computed by implementing the SR and curved space-time transformations on the specific disc intensities and directional derivatives. And as expected, the curvature in space reduces the magnitude of the radiative moments. However, the effect of radiation is more complicated than what meets the eye. Radiation
drag term, when computed in GR regime, overwhelms near the horizon because of the presence of $1/g^{rr}$ term,
compared to flat space. But it is lesser than that computed in flat space-time, further out. Crucially, this departure of computing drag term in GR from flat space value cannot be mimicked by some simple scaling relation.

In the advective disc model, there are two sources of radiation --- the inner compact corona and the outer disc.
The accretion rate not only controls the overall radiative output from the disc, but also determines the
size of the corona. 
Since we are considering Thomson scattering regime, the details of the
spectrum do not matter and frequency integrated moments of the radiation field suffice.
The radiative moments generally have two peaks corresponding to the radiation from the corona and the outer disc
(Figs. \ref{lab:fig2}). A comparison of the moments for an accretion disc
with an inner corona and outer KD \citep{cdc04, c05} with the present disc model shows that the radiative moment computed
from the outer disc of the present model are much stronger.

In this paper, we computed the generalized, relativistic Bernoulli parameter ($E$) for radiatively driven flow in curved space time. This is a constant of motion even in the presence of radiation driving. The expression of relativistic
Bernoulli parameter ($\equiv -hu_t$) for adiabatic and isentropic flow is not conserved along the streamline of a radiatively driven flow, or across the shock but, $E$ is a constant of motion. This gives us a great tool to find various classes of solutions. One should not be
confuse $E$ with the generalized relativistic Bernoulli parameter obtained for accretion discs \citep{ck16,kc17}.
Since the streamline and various dissipative processes in an accretion disc are different than the jet
(compare $X_f$ of equation \ref{energy.eq} of this paper and equation 18 of Chattopadhyay \& Kumar 2016),
the values of generalized Bernoulli parameters will not be the same for jet and accretion disc,
even if the jet is launched with the local accretion disc variables on the foot points of the jet. 
 
In this paper, unlike \citep{vkmc15}, we considered hotter and therefore geometrically thicker corona.
This has a very interesting radiative flux ($R_1$) distribution. Within the funnel of the corona,
$R_1<0$ and therefore opposes the out-flowing jet. Above the height of the corona, $R_1>0$ and it pushes the jet outward. That the radiation accelerates, can be
understood from the fact that the range of sonic point gets limited, with the increase of disc luminosity.
If $E$ is high, then the jet is hot at the base
and the effect of radiation is negligible. Thermal driving completely dominates within the funnel
and accelerates the jet such that $v\sim \veq$.
Above the funnel the jet is sufficiently cooled, such that the radiative term starts to become effective, but since the jet has reached up to the local equilibrium speed, radiation deceleration would actually slow the jet down (Figs. \ref{lab:fig9n}b, \ref{lab:fig10n}a). For medium and small values of $E$, thermal and radiation
driving may accelerate jets to relativistic speeds and the speed increases with the disc luminosity.
In fact, for lepton dominated flow ($\xi=0.01$) jets
do reach $\gamt \gsim 10$. But more than acting just as an agent of acceleration/deceleration,
radiation does trigger
a shock transition in jets very close to the BH. The shock range is small and the shock strength
is moderate and peaks at certain values of jet energy for a given disc luminosity.
It may be noted that, shocks generated in this paper are triggered by the inwardly directed radiation
flux within the funnel of the corona, which is different than the shocks generated by `pinching off' the flow geometry in \cite{vc17}.

Radiatively driven fluid jet in relativity, has a very rich class of solutions. The `e' type solutions 
may have one inner type sonic point, multiple sonic points and shocks. While the `f' type
jet is a low energy solution, such solutions passes through the outer sonic point.
The radiative driving is the most effective for `f'-type jet solutions (Fig. \ref{lab:fig12n}a). This class of solutions can be compared with radiatively driven $\el$ jets in the particle approximation
\citep{cdc04,c05}. Interestingly, discs with sub-Eddington luminosity can power lepton dominated
jets ($\xi=0.01$) to terminal Lorentz factors $\gamt \sim 3$, but super-Eddington discs can power those
f-type jets to $\gamt \sim 10$ (Fig. \ref{lab:fig12n}b). We have earlier
argued that the radiation driving of particle jets, is more efficient than the fluid one because of the presence of
the enthalpy term in the denominator of radiation term (equation \ref{radterm.eq}). However, the advantage
of considering radiation driving of fluid jets is that, where ever the jet has been hot, radiation driving
is not effective,
but the thermal gradient term is. In the region where, the temperature falls down, thermal gradient becomes less effective, but radiation takes over, provided the region is relatively closer to the disc ($\sim 100 \rg$).
Therefore, the lepton dominated jets achieve terminal speeds similar to the $\el$ particle jets,
in addition, the radiation driving can produce fluid phenomena like shocks in the jet. An unstable
shock can also produce effects like QPOs in the jet, a scenario worth investigating.
Moreover, such internal shocks close to the jet base have been invoked to explain the
high energy power-law tails in some of the microquasars \citep{l11}.
\cite{ftrt85} also showed the existence of shocks in radiatively driven jets, when the disc was quite
thick and jet geometry deviates from the conical geometry. Although the authors were not considering the
effect of acceleration of radiation on jets, but nonetheless, the $\vt$ quoted by them were all mildly relativistic
($\vt \sim 0.1$). Whereas, in our paper, we find the $\vt$ is few times higher in general. The reason being
that \cite{ftrt85} considered mostly isothermal jets and therefore missed the thermal driving factor for
the jet. 
Our present work is also different from \cite{mstv04} since the accelerating agent in their
work was hidden within the equation of state. They also did not find any fluid discontinuities like shock in the jets.

We would conclude by stating that, radiation is an important agent in triggering various physical processes
in a jet. The radiation can drive $\ep$ jets to reasonable terminal speeds ($\vt \gsim 0.5$) if the disc is
sub-Eddington. However, for very hot jets under intense radiation field, $\gamt\sim 3$ is achievable.
For lepton dominated flow and intense radiation field $\gamt \sim 10$ is also possible. The response of jet terminal speed with
disc luminosity is not straight forward, $\vt$ may slightly decrease with increasing luminosity
for high energy jet, it may decrease and then increase with increasing luminosity for moderate energy jets,
but will increase with $\ell$ for low energy jets. It may be worth noting that radiation may accelerate
jets to relativistic terminal speeds, contrary to what is popularly accepted \citep{ggmm02}. 

%{\textbf{\color{blue} Quoting the significance of f-type solutions obtained in this paper, we revert the previously established assumption that radiation is not able to drive jets up to relativistic speeds if the base is not hot enough \citep{ggmm02}.
%In addition, in the very first effort we show that radiation field alone can generate internal shocks close to the BH. Internal shocks close to the BH are good sites for particle acceleration and are required to explain high energy emissions from such sources \citep{l11}.}} 
\section*{Acknowledgment} 
The authors acknowledge the anonymous referee for raising pertinent issues which helped improving the
quality of the paper. The authors also acknowledge ARIES for supporting this work.

\appendix
\section{Accretion Disc and associated radiation parameters}
\subsection{Estimating approximate accretion disc variables}
%\subsubsection{Estimating approximate accretion disc variables}
$U^\mu$ are the components of accretion
four-velocity, and the corresponding three-velocity components are
${\bf v}\equiv(\vartheta_x,0,\vartheta_\phi)$, where $x,~\theta,~\phi$ are usual spatial coordinates. We define
$\vartheta=\vartheta_x/\sqrt{(1-\vartheta_\phi^2)}$ as the radial three-velocity measured
by a local rotating observer. Following this, one can present
the velocity distribution of the outer disc and the corona in a compact form
\citep[see Appendix A of][]{vkmc15}
\be
\vartheta_{\rm i}=\left[1-\frac{(x-2)x^2}{\{x^3-[(x-2)\lambda^2]\}U_t^2|_{x_{0\rm i}}}
\right]^{1/2}.
\label{accvel.eq}
\ee
Here, the suffix ${\rm i}$ denotes variables of the corona (i. e.,
i$=${\small C}) or the outer disc (i. e., i$=${\small D}) and
$U_t|_{x_{0\rm i}}$ is the covariant time component of the $U^\mu$s at the outer edge.
For the corona, $x_{0\rm i}=\xsh$ and for the outer disc $x_{0\rm i}=x_0$.
At $x_0$, $ [\vartheta_{\rm \small D}]_{x_0} \approx 0$ but increases as it falls towards the BH till it reaches
$\xsh$, where the flow speed reduces by one-third. In shocked accretion disc, this reduction is automatic, but
even in shock free discs centrifugal barrier, radiation pressure all can impede the inflow, making it hot and thereby forming the corona. 
Assuming a slow variation of the adiabatic index the temperature distribution can also be assumed as \citep{vkmc15}
\be
\Theta_{\rm i}=\Theta_0\left(\frac{U^x_0x_0H_0}{U^x_{\rm i} xH_{\rm i}}\right)^{\Gamma -1}.
\label{acctemp.eq}
\ee
Moreover, \cite{vkmc15} proposed an approximate relation between $\xsh$ and the
accretion rate, given by
%\bea
%\xsh=125.31326199423765-24.602485347397550{\dot m} \nonumber \\
%+1.7646624611542214{\dot m}^2-0.043447536123535316{\dot m}^3
%\label{xsdotm.eq}
%\eea
\be
\xsh=125.313-24.603{\dot m}+1.765{\dot m}^2-0.043{\dot m}^3
\label{xsdotm.eq}
\ee
Here $\xsh$ is in geometric units and $\dot m$ is the accretion rate in units of Eddington
rate (Eddington rate $\equiv {\dot M}_{\rm Edd}=1.4\times10^{17}\mbh/\msol$gs$^{-1}$).
In order to completely specify $\vartheta_{\rm i}$ and $\Theta_{\rm i}$ at all $x$, one also needs to know the local height $H_{\rm i}$.
Numerical simulations show that the outer disc has a flatter structure than that predicted by
assumptions of vertical equilibrium and the inner torus like corona is basically a thick
disc (with advection terms) and the height to radius ratio can vary anything between 1.5 to 10  \citep{dcnm14,lckhr16}. Therefore, we define $H_{\rm 0}=0.4H_{\rm sh}+\tan \theta_{\rm D}x_0$.
If we supply $[\vartheta_{\rm \small D}]_{x_0},~\rho_0,~H_0$ and ${\dot m}$ at $x_0$, then the distribution
of velocity, temperature, density at all $x_{\rm i}$ and the location of $\xsh$ can be estimated. Typical accretion disc parameters
are given in table \ref{table1}.
\subsection{Radiative intensity and luminosity from the accretion flow}
The outer disc emits mainly via synchrotron and bremsstrahlung processes and the corona
additionally via inverse-Compton process. 
The functional form of the frequency integrated, local intensity of the outer disc is given by \citep{kc14,vkmc15}, 
$$
{\tilde I}_{\od}={\tilde I}_{\rm syn}+{\tilde I}_{\rm brem}
$$
\bea
=\left[\frac{16}{3}\frac{e^2}{c}\left( \frac{eB_{\od}}{m_e c} \right)^2
 \Theta^2_{\od} n_{\od} x+ 1.4\times 10^{-27}n_{\od}^2g_bc \sqrt{\frac{\Theta_{\od} 
m_e}{k}}\right] \nonumber \\
\times \frac{\left(d_0 \sin \theta_{\od}+x\cos \theta_{\od}
\right)}{3} \ \  {\rm erg} \
{\rm cm}^{-2} {\rm s}^{-1}
\label{skint.eq}
\eea
Here, $\Theta_{\od}, n_{\od}, x, \theta_{\od}$, $B_{\od}$ and $g_b(=1+1.78\Theta_{\od}^{1.34})$ are the 
local dimensionless temperature, electron number density, radial distance of the disc, the semi-vertical
angle of the outer disc surface, 
the magnetic field and relativistic Gaunt factor, respectively. Intensity is measured in the disc local rest
frame.
The factor outside square brackets
converts emissivity (${\rm erg}~{\rm cm}^{-3}{s}^{-1}$) into intensity  (${\rm erg}~ {\rm cm}^{-2}{s}^{-1}$).
The luminosity of the outer disc is obtained by integrating $I_0$ over the disc surface, {\ie}
\be
L_{\od}=2\int^{x_{0}}_{\xsh} \int^{2\pi}_{0} I_{\od}r~{\rm cosec}^2\theta_{\od}~d\phi dx
\label{lum.eq}
\ee
which, we can be presented in units of $L_{\rm Edd}(\equiv 1.38\times10^{38}\mbh/\msol~{\rm ergs}{\rm s}^{-1})$ as $\ell_{\od}=L_{\od}/L_{\rm Edd}$.
Since the accretion disc solution has been approximated, so
we do not calculate the radiation from corona directly, but instead estimate it from the enhancement
factor computed from self-consistent two temperature solutions \citep{mc08}. 
The ratio of corona and outer disc luminosities, is computed following \cite{mc08} and was presented
in \cite{vkmc15},
\begin{eqnarray}
\chi=-5.974+1.996 \xsh -0.166 \xsh^2 + 6.653 \times 10^{-3} \xsh^3 \nonumber \\
   -1.280 \times 10^{-4} \xsh^4 + 9.455\times 10^{-7} \xsh^5~(\xsh < 35); \nonumber \\
\chi=2.693+0.096 \xsh-3.465 \times 10^{-3} \xsh^2 + 3.898\times 10^{-5} \xsh^3  \nonumber \\
-1.439\times 10^{-7} \xsh^4 ~(\xsh \geq 35)% \nonumber
\label{chi1.eq}
\end{eqnarray}
We assume that these functions are generic.
So the luminosity from the corona can be estimated as $L_{\rm \small C}= \chi L_{\od}$,
and the dimensionless total luminosity is given by
\begin{eqnarray}
\ell=\ell_{\rm \small C}+\ell_{\rm \small D}=(1+\chi)\ell_{\rm \small D}
%;~~\mbox{for}\mbh=\Ma \nonumber \\
%~~~  =\chi_8 L_{\od};~~\mbox{for}\mbh=\Mb
\label{ell.eq}
%\label{coronalum.eq}
\end{eqnarray}
%The total disc luminosity $\ell$ (in units of $L_{\rm Edd}$) is given by,
%\be
%%\ell=\ell_{\rm \small C}+\ell_{\rm \small D}=(1+\chi)\ell_{\rm \small D}
%
%\ee
The specific intensity measured in the local rest frame of the corona is given by 
${\tilde I}_{\rm \small C}= L_C/{\pi}{A_C}$. The dimensionless form of $\sigma_T {\tilde I}_{\rm \small C}/m_e$
is given by
\be
\frac{\sigma_T{\tilde I}_{\rm \small C}}{m_e}=\frac{1.3{\times}10^{38}{\ell}_{\od} \chi \sigma_T}{2{\pi}c m_e{A}_{\rm \small C} GM_{\odot}} 
\label{coronaint.eq}
\ee
Here $A_{\rm \small C}$
is the surface area of the corona. 
To obtain the specific radiation intensities (equations \ref{skint.eq} and \ref{coronaint.eq}) from the accretion disc, we need the number density and temperature distribution of the disc. Here, ${\tilde I}_{\rm \small C}$ is obtained from ${\tilde I}_{\od}$, in which $n_{\rm \small D}$ 
is obtained by supplying ${\dot m} $ and equation(\ref{accvel.eq}), and $\Theta_{\rm \small D}$
from equation(\ref{acctemp.eq}). 
In this paper, we have only concentrated on accretion discs around $\mbh=10\msol$.
\begin{table}
\caption{Disc parameters}
\label{table1}
\centering
 \begin{tabular}{|c c c c c c c|} 
 \hline
 $\lambda$ & $x_0$ & $\left[\vartheta_{\rm \small D}\right]_{x_0}$ & $\left[\Theta_{\rm \small D}\right]_{x_0}$ & $\theta_D$ & $H_{\rm sh}$ & $d_0$\\ [0.5ex] 
 \hline%\hline
 $1.7$ & $5500 \rs$ & $1.5 \times 10^{-3}$ & $0.2$ & $85^0$ & $2.5\xsh$ & $0.4H_{\rm sh}$\\ 
 \hline
 \end{tabular}
\end{table}
%\end{center} 
%\section*{APPENDIX B : Paczy\'nski-Wiita potential and general relativity}
\section{Paczy\'nski-Wiita potential (PW) and relativistic flows}
More often, radiatively driven relativistic jets are studied in the SR plus PW regime and not in GR regime.
Although Equivalence principle strictly precludes this possibility, but in astrophysics this trend has been 
followed by a number of researchers because it is assumed that GR affects only in the region outside the
BH and not at moderate to large distances. Here we show that, the differences in equations in the two approach
affects the solutions close to the BH, as well as at moderate distances (few$\times~10\rg$). Moreover, the temperature produced in SR+PW regime is unphysically high. Furthermore, radiation effects in curved space time
cannot be properly taken into account by any scaling relations, if the moments are computed in the flat space.
The effect of curvature in the radiation term has been addressed in section \ref{sec:rad_curved}.
So we list the first two points below. 
\subsection{Equations of motion}
%\textbf{(i) Paczynski-Wiita (PW) potential vs GR calculations}\\
The equations of motion in the two approaches, can be written down as,
%\begin{equation}
%\left[\frac{dv}{dr}\right]_{GR}=\frac{2a^2r-{\color{red} \frac{\left(1-a^2\right)r}{r-2}}+\frac{(2-\xi)\gamma r^2}{(f+2\Theta){\color{red} \sqrt{g^{rr}}}}\left[(1+v^2){\color{red}{\cal R}_1}-v
%{\color{red}\left(g^{rr} {\cal R}_0+\frac{{\cal R}_2}{g^{rr}}\right)}\right]}{r^2\gamma^2v\left(1-\frac{a^2}{v^2}\right)}
%\end{equation}
\begin{equation}
\begin{split}
\gamma^4v&\left(1-\frac{a^2}{v^2}\right)\left[\frac{dv}{dr}\right]_{GR}=\frac{2a^2 \gamma^2}{r}-{\frac{\left(1-a^2\right)\gamma^2}{r(r-2)}} \\
& +\frac{(2-\xi)\gamma^3}{(f+2\Theta){ \sqrt{g^{rr}}}}\left[(1+v^2){{\cal R}_1}-v
{\left(g^{rr} {\cal R}_0+\frac{{\cal R}_2}{g^{rr}}\right)}\right]
\end{split}
\label{appb1.eq}
\end{equation}
and
%\begin{equation}
%\left[\frac{dv}{dr}\right]_{PW}=\frac{2a^2r-{\color{red} \frac{(1-v^2)r^2}{{(r-2)^2}}}+\frac{(2-\xi)\gamma r^2}{(f+2\Theta)}\left[(1+v^2){\color{red}{\cal R}_{1F}}-v
%{\color{red} \left({\cal R}_{0F}+{\cal R}_{2F}\right)}\right]}{r^2\gamma^2v\left(1-\frac{a^2}{v^2}\right)}
%\end{equation}
\begin{equation}
\begin{split}
\gamma^4v&\left(1-\frac{a^2}{v^2}\right)\left[\frac{dv}{dr}\right]_{PW}=\frac{2a^2 \gamma^2}{r}-{\frac{1}{{(r-2)^2}}} \\
&+\frac{(2-\xi)\gamma^3}{(f+2\Theta)}\left[(1+v^2){{\cal R}_{1F}}-v
{ \left({\cal R}_{0F}+{\cal R}_{2F}\right)}\right]
\end{split}
\label{appb2.eq}
\end{equation}
The subscript PW signifies equations of motion in SR+PW regime, while GR represent the equations in
Schwarzschild metric.
Values with subscript $F$ are calculated in flat space.
The r. h. s of the two equations (\ref{appb1.eq} and \ref{appb2.eq}) algebraically differ in the second term of the numerator,
and the presence of curvature $g^{rr}$ in the radiation terms. 
Since the form of the first terms are same, let us take the ratio of the 2$^{nd}$
terms on r.h.s of the above two equations,
$$
\frac{(1-a^2)g^{rr}}{{(1-v^2)}}=\delta
$$
The calculations with SR+PW potential will be comparable to GR if $\delta\sim1$. 
%This condition reveals that PW potential has been successful in solving accretion problems but will give large error in study of relativistic outflows. In spherical accretion, at larger distances, both $v$ and $a$ are much smaller than 1 and as $g^{rr}\rightarrow1$, this condition is satisfied by remarkable precision. As fluid comes closer to the BH, $(1-v^2)$ decreases faster than $(1-a^2)$ and it is compensated by decrease in $g^{rr}$. On the contrary
For relativistic winds or jets, at large distances, $g^{rr}\approx 1$, $a$ is very low but $v\rightarrow 1$, then
$$
\delta \gg 1
$$
Similarly close to the BH i.e., $r\rightarrow 2$, $g^{rr} \rightarrow 0$,
$v\approx 0$, but $a$ is large, therefore
$$
\delta \ll 1
$$
This analysis points to the possibility
of a large deviation from GR solutions even at larger distances. \\
This difference in the second term of the EoMs arise because in SR+PW, regime the gravity enters as an additive
term (equation \ref{appb2.eq}), while in GR it is a space time phenomena (equation \ref{appb1.eq})
so it affects any source of energy. Therefore, the curvature term ($g^{\mu \nu}$s) should couple with the thermal and the kinetic
terms as is seen above, i. e., the curvature term is coupled with the thermal term in the form of sound speed $a$
and also the Lorentz factor $\gamma$. Consequently, if there is a discontinuity like shock in the flow,
then the gravity term in SR+PW will not change across the shock, but in GR it will change, since
both $a$ and $v$ jump across a shock.
If one may add further,
\cite{abci96} also showed that even in accretion problems, SR and PW potential are not compatible.
\subsection{Overestimated thermal content in PW analysis}
\begin{figure}[h]
\begin{center}
 \includegraphics[width=8.cm, angle=0, trim=0 0 160 180,clip]{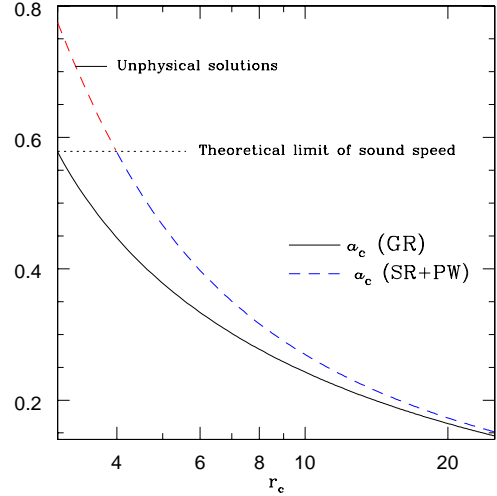}
\vskip -0.5cm
 \caption{$a_c$ as a function of $r_c$ for $\ell=0$. % Right : $\Theta_c$ as a function of $r_c$ for $\ell=2.85$. 
 Solid curve is for GR solutions while dotted shows solutions with PW potential
}
\label{lab:figB1}
 \end{center}
\end{figure}
Causality imposes an upper limit of sound speed which is $a<1/\sqrt{3}$. In GR this translates to a lower bound
in the location of sonic points ($r_c>3$). This is clear in Figs. (\ref{lab:fig5n}). Since pseudo potentials makes the flow unphysically hot, so the lower limit
of sonic point in pNp+SR regime extends to a larger distance ($r_c>4$).  
For thermally driven flow this can be very easily shown. From equations (\ref{appb1.eq}) and (\ref{appb2.eq})
and ignoring radiation, we obtain $a_c$ as a function of $r_c$. 
$$
[a_c]^2	_{GR}=\frac{1}{2r_c-3}
$$
$$
[a_c]^2	_{PW}=\frac{r_c}{2(r_c-2)^2+r_c}
$$
These are algebraic relations, and $[a_c]_{PW}$
is higher than $[a_c]_{GR}$ (Fig. \ref{lab:figB1}). In other words, the jet in SR+PW description
is much hotter than the GR one. This also means the SR+PW jet is subjected to a much stronger
thermal gradient push than it happens in reality. Moreover, all jet solutions with $3<r_c\leq 4$ are absent in
SR+PW solutions.

\end{document}